%% file: ms.tex
\documentclass[structabstract]{aa}  
\usepackage{graphicx}
\usepackage{txfonts}
\usepackage{natbib}
\def\LSXI{\mbox{LS~III~+46~11}}
\def\LSXII{\mbox{LS~III~+46~12}}
\def\BXC{\mbox{Berkeley~90}}
\def\teff{\mbox{$T_{\rm eff}$}}
\def\logg{\mbox{$\log g$}}
\def\vsini{\mbox{$v\sin i$}}
\def\ebv{\mbox{$E(4405-5495)$}}
\def\rv{\mbox{$R_{5495}$}}
\def\logd{\mbox{$\log d$}}

\def\AV{\mbox{$A_V$}}
\def\mum1{\mbox{$\mu$m$^{-1}$}}
\def\chir{\mbox{$\chi^2_{\rm red}$}}

\newcommand{\HeII}[1]{\mbox{He\,{\sc ii}~$\lambda${#1}}}

\newcommand{\CH}[1]{\mbox{CH~$\lambda${#1}}}
\newcommand{\CHp}[1]{\mbox{CH+~$\lambda${#1}}}
\newcommand{\CN}[1]{\mbox{CN~$\lambda${#1}}}
\newcommand{\NaI}[1]{\mbox{Na\,{\sc i}~$\lambda${#1}}}
\newcommand{\CaI}[1]{\mbox{Ca\,{\sc i}~$\lambda${#1}}}
\newcommand{\CaII}[1]{\mbox{Ca\,{\sc ii}~$\lambda${#1}}}
\newcommand{\KI}[1]{\mbox{K\,{\sc i}~$\lambda${#1}}}
\newcommand{\DIB}[1]{\mbox{DIB~$\lambda${#1}}}
\begin{document}
   \title{The little-studied cluster Berkeley 90}
   \subtitle{II. The foreground ISM}


   \author{J. Ma{\'\i}z Apell{\'a}niz\inst{1}
        	  \and
	          R. H. Barb\'a\inst{2}
	          \and
	          A. Sota\inst{3}
	          \and
           S. Sim\'on-D{\'\i}az\inst{4,5}
          }

   \institute{Centro de Astrobiolog{\'\i}a, CSIC-INTA, campus ESAC, apartado postal 78, E-28\,691 Villanueva de la Ca\~nada, Madrid, Spain \\
	      \email{jmaiz@cab.inta-csic.es} \\
         \and
              Departamento de F{\'\i}sica y Astronom{\'\i}a, Universidad de La Serena, Av. Cisternas 1200 Norte, La Serena, Chile \\
         \and
              Instituto de Astrof{\'\i}sica de Andaluc{\'\i}a-CSIC, Glorieta de la Astronom\'{\i}a s/n, E-18\,008 Granada, Spain \\
         \and
              Instituto de Astrof{\'\i}sica de Canarias, E-38\,200 La Laguna, Tenerife, Spain \\
         \and
              Departamento de Astrof{\'\i}sica, Universidad de La Laguna, E-38\,205 La Laguna, Tenerife, Spain \\
             }

   \date{Received XX XXX 2015; accepted XX XXX 2015}

 
  \abstract
  {Nearly one century after their discovery, the carrier(s) of Diffuse Interstellar Bands is/are still unknown and there are few sightlines studied in detail for a 
   large number of DIBs.}
  {We want to study the ISM sightlines towards \LSXI\ and \LSXII, two early-O-type stellar systems, and \LSXI~B, a mid-B-type star. The three targets are located 
   in the stellar cluster \BXC\ and have a high extinction.}
   {We use the multi-epoch high-S/N optical spectra presented in Paper I (Ma{\'\i}z Apell{\'a}niz et al. 2015), the extinction results derived there, and additional
  spectra.}
   {We have measured equivalent widths, velocities, and FWHMs for a large number of absorption lines in the rich ISM spectrum in front of \BXC.
    The absorbing ISM has at least two clouds at different velocities, one with a lower column density (thinner) in the K\,{\sc i} lines located 
    away from \BXC\ and another one with a higher column density (thicker) associated with the cluster. The first cloud has similar properties for both O-star 
    sightlines but the second one is thicker for \LSXI. The comparison between species indicate that the cloud with a higher column density 
    has a denser core, 
    allowing us to classify the DIBs in a $\sigma-\zeta$ scale, some of them for the first time. The \LSXII\ sightline also has a high-velocity redshifted component.}
   {}


   \maketitle
%

\section{Introduction}

$\,\!$ \indent Diffuse interstellar bands (DIBs) were discovered nearly a century ago \citep{Hege22,Merr34}. There are many of them known, with FWHMs that range from 
under 1~\AA\ to several tens of \AA. They are correlated with the amount of extinction but not perfectly so. To date, the carrier or carriers that produce
them have not been identified with certainty. Different options have been proposed, most of them carbon-based substances 
\citep{Herb95,Ehreetal95,Thoretal03,Kazmetal10,Maieetal11,Salaetal11,Stegetal11}.

Broadly speaking, DIB observational studies can be classified in two categories. Those in the first one select one or a small number of DIBs and measure them for
a large sample of stars. The pioneering study was \citet{Duke51} and in recent years this type has become popular
\citep{Thoretal03,Munaetal08,Frieetal11,Vosetal11,Raimetal12,Puspetal13,vanLetal13}. The second category corresponds to studies that select one or a small number of 
stars and study their DIBs in depth throughout a large wavelength range \citep{JennDese94,Galaetal00,Tuaietal00,Coxetal05,Hobbetal08,Hobbetal09}. The work presented 
here is of the second category, though it is part of a larger project that plans to study the ISM using thousands of high-quality spectra of OB stars 
\citep{Penaetal13,Maizetal14b}.

\begin{figure*}
\centerline{\includegraphics[width=0.49\linewidth]{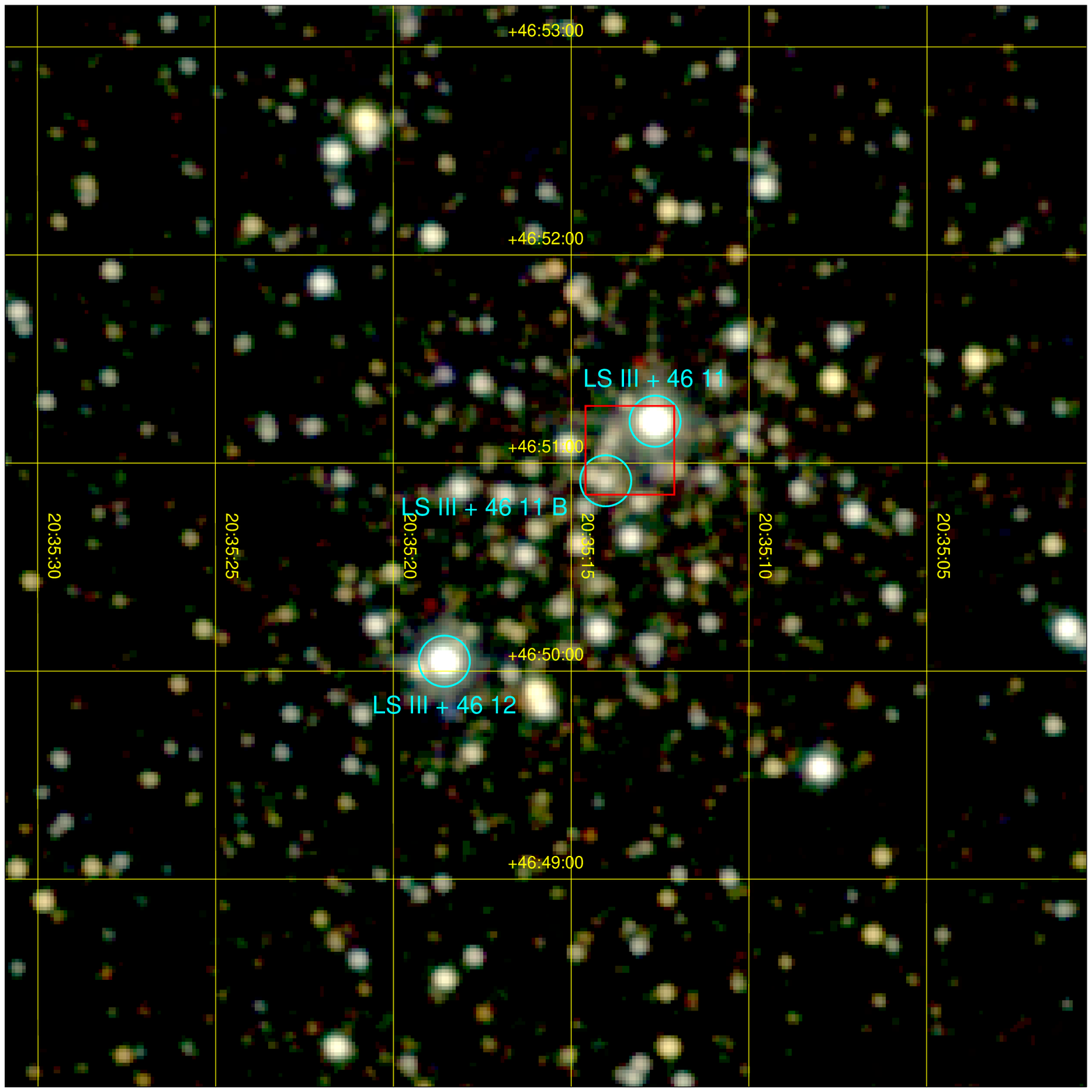} \
            \includegraphics[width=0.49\linewidth]{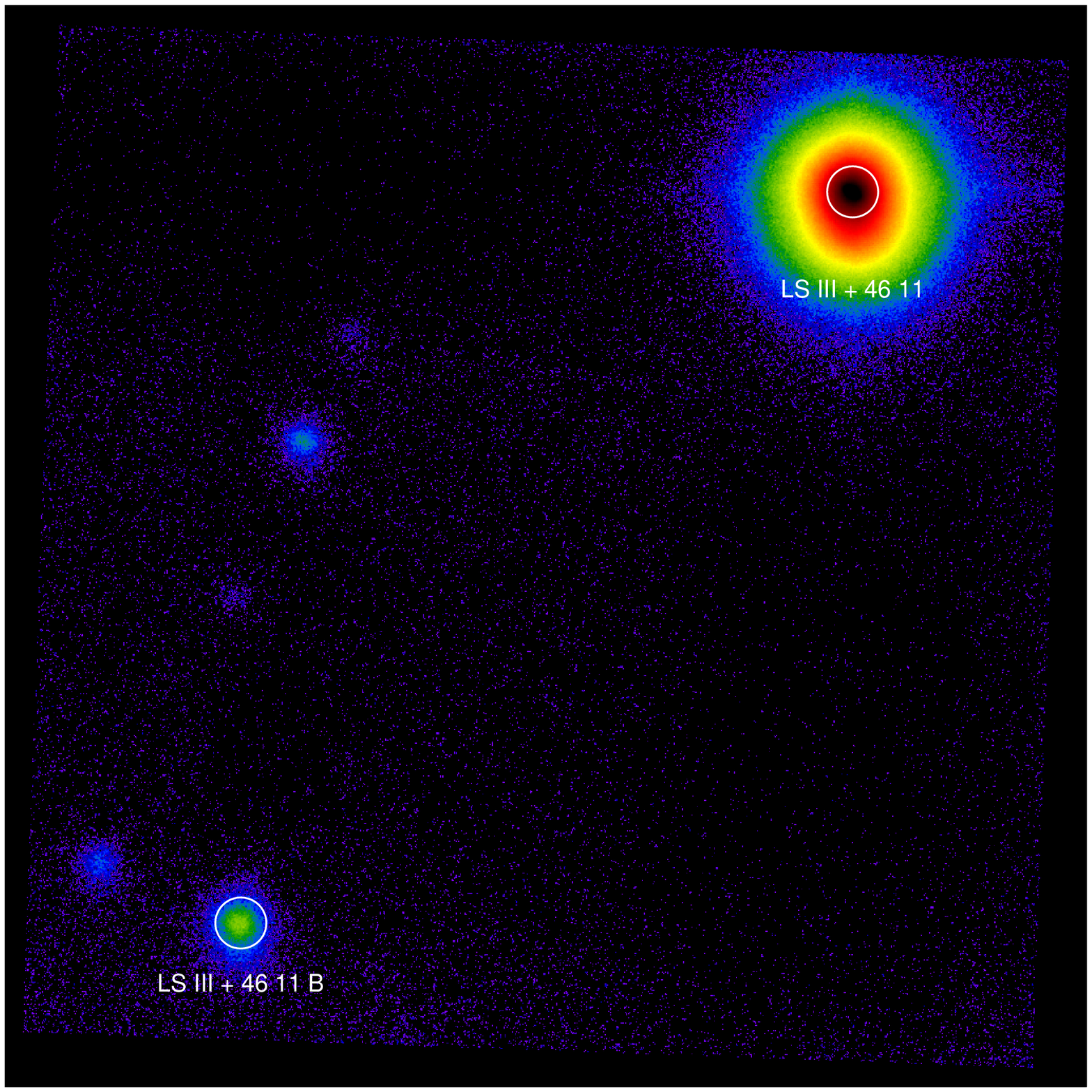}}
\caption{[left] 2MASS $K_{\mathrm{S}}HJ$ three-color RGB mosaic of \BXC. The intensity level in each channel is logarithmic. [right] False-color AstraLux 
         $25\farcs 6\times 25\farcs 6$ $i$-band image of \LSXI\ (see Paper I). North is up and East is left and the image location corresponds to the red square in the
         left panel. The intensity level is logarithmic.}
\label{images}
\end{figure*}

\begin{table*}
\centerline{
\begin{tabular}{lcccl}
                & \LSXI\              & \LSXII\          & \LSXI~B                    & Reference                   \\
\hline
Sp. type        & O3.5~If*~+~O3.5~If* & O4.5~V((f))      & B4~V                       & Paper I, this work          \\
RA (J2000)      & 20:35:12.642        & 20:35:18.566     & 20:35:14.029               & \citet{Hogetal00a}          \\
dec (J2000)     & +46:51:12.12        & +46:50:02.90     & +46:50:54.94               & \citet{Hogetal00a}          \\
$l$ (deg.)      & 84.8844             & 84.8791          & 84.8830                    & \citet{Hogetal00a}          \\
$b$ (deg.)      & +3.8086             & +3.7836          & +3.8026                    & \citet{Hogetal00a}          \\
$V_{J}$         & 10.889$\pm$0.021    & 10.268$\pm$0.009 & $\;\,$15.992$\pm$0.033$^*$ & Paper I, this work          \\
$K_{\mathrm S}$ & 6.971$\pm$0.023     & 7.470$\pm$0.023  & 11.821$\pm$0.053           & \citet{Skruetal06}, Paper I \\
\rv             & 3.303$\pm$0.058     & 3.377$\pm$0.040  & 3.34 (fixed)               & Paper I, this work          \\
\ebv\           & 1.653$\pm$0.020     & 1.255$\pm$0.011  & 1.590$\pm$0.019            & Paper I, this work          \\
\AV\            & 5.475$\pm$0.037     & 4.272$\pm$0.021  & 5.324$\pm$0.064            & Paper I, this work          \\
$v$ (km/s)      & $-$20$\pm$5         & $-$13$\pm$3      & ---                        & Paper I                     \\
\hline
\multicolumn{5}{l}{$^*$ Calculated from CHORIZOS, not measured.} \\
\end{tabular}
}
\caption{\LSXI, \LSXII, and \LSXI~B summary.}
\label{summary}
\end{table*}

The sparse young open cluster \BXC\ is almost unstudied. In \citet{Maizetal15a}, from now on Paper I, we analyzed it with an emphasis on
\LSXI, a very massive early-type binary composed of two of the four O stars earlier than O4 known in the northern hemisphere. Paper I also studied
\LSXII, the other early-O-type system with a significant contribution to the ionizing flux of the cluster, and other properties of \BXC. The cluster is immersed 
in an H\,{\sc ii} region, Sh 2-115. In Table~\ref{summary} we summarize some of the properties of the two dominant systems in the cluster 
and of a third star that is also studied here.
Note that the two dominant stars
experience a similar extinction law, as evidenced by the measured \rv\ values \citep{Maiz13b,Maizetal14a}, but \LSXI\ has an extinction which is higher by 
$\approx 30\%$. In this paper we study the signature of the foreground ISM in the optical spectra of both stars. We also include in our study a third system, \LSXI~B, 
which is spatially resolved from \LSXI\ (Fig.~\ref{images}) and whose photometry was presented in Paper I, for which we have also obtained spectra.

\section{Data}

$\,\!$ \indent Most of the spectra used here are a selection of those used in Paper I. The reader is referred to that article for details; here we only summarize 
the most relevant information. The spectra were obtained under four different projects (GOSSS, \citealt{Maizetal11}; NoMaDS, \citealt{Maizetal12}, \citealt{Pelletal12};
IACOB, \citealt{SimDetal11c}, \citealt{SimDetal15b}; and CAF\'E-BEANS, \citealt{Neguetal15}) at resolving powers ranging from 2800 to 65\,000 and covering different
ranges from 3811~\AA\ to 9225~\AA. 

In addition to the spectra from Paper I, we have obtained long-slit spectroscopy on 29 April 2015 with the OSIRIS instrument at the 10.4 m Gran Telescopio Canarias 
(GTC), as part of an extension of GOSSS to larger-aperture telescopes to include fainter O stars. Two gratings were used, R2500U (3440-4610~\AA, $t_{\rm exp}$ of 1800~s) 
and R2500V (4445-6060~\AA, $t_{\rm exp}$ of 300~s), with a resolving power $R$ between 2500 and 3000. As in the rest of the GOSSS data, the long slit was placed to 
include both \LSXI\ and \LSXII. The larger aperture of GTC allowed us to extract the spectra of a third star, \LSXI~B (2MASS~J20351402+4650549, see Paper I and 
Fig.~\ref{images}), that was present in the long-slit exposures.

Note that the number of epochs, S/N, and wavelength coverage at each resolution for each star varies, so there are differences in the ISM lines that can be 
studied for each star. In particular, the blue-violet range was significantly less covered for \LSXII\ than for \LSXI\ and the number of lines that could be measured for
\LSXI~B was relatively small. The telluric lines in the high-resolution spectra were eliminated according to the procedure of \citet{Gardetal13}. 

The results in this paper can be divided into two types. We first conduct a brief analysis of the spectra and photometry of \LSXI~B in order to derive its properties 
and place it in the context of \BXC. We then study the ISM in front of the three targets, which is the core of this paper.

\input{ism1}

For the ISM we have measured as many as seven atomic and seven molecular lines and fifty DIBs. We give the equivalent widths (EWs), full-widths at half maximum (FWHMs), 
and central velocities ($v$) for \LSXI\ and \LSXII\ obtained with the WHT and the high-resolution spectroscopy in Table~\ref{ism1}. Before analyzing the results, we 
detail some aspects of the extraction:

\begin{itemize}
 \item Our spectra are diverse in terms of wavelength coverage, S/N, and spectral resolution. Therefore, each absorption line was extracted spectrum by spectrum 
       and the results were combined a posteriori using the best available data in each case.
 \item In cases where spectra of similar resolutions but different S/N values were available, we only used those where the S/N was adequate.
 \item For narrow lines, high-resolution spectroscopy is preferred over intermediate-resolution data. For broad DIBs, however, the situation is reversed as 
       measurement errors are dominated by normalization, which is easier to carry out in long-slit spectra than in echelle data. For the intermediate-resolution 
       case we only used the WHT spectra due to their slightly higher spectral resolution compared to the CAHA-3.5~m data.
 \item In most cases, fitting with a single Gaussian yields better results than numerical integration and the former are the ones displayed in Fig.~\ref{ism1}. 
       However, some lines are asymmetric due to their intrinsic profiles \citep{Galaetal08} or the existence of two or three kinematic 
       components (see below). For those lines we display the numerical integration results in Fig.~\ref{ism1}. 
 \item No attempt was done to fit broad DIBs with Lorentzian profiles instead of Gaussians \citep{Snowetal02,vanLetal13} because of the rectification difficulties 
       that exist for the wings. Note, however, that differences in the results are small \citep{Maizetal14b}.
 \item For the DIB groups at 4761.12+4762.36+4779.69~\AA\ and 4879.83+4887.43~\AA\ \citep{Maizetal14b}, the lines were fitted simultaneously. Note, however, that for 
       the DIB at 6203.05~\AA\ only the inner narrow DIB was fit (but not the broad one) and that a single Gaussian was also used for the 6177.30~\AA~DIB despite its
       obvious asymmetry due to the complexity of the spectral region\footnote{
       The velocities for the 6177.30~\AA~DIB differ from the rest of the DIBs precisely for this reason and should not be considered; only its EWs in Table~\ref{ism1}
       are of practical use. Nevertheless, note that velocities for \LSXI\ and \LSXII\ are consistent between them, a sign that they originate in the same clouds.
       }.
 \item The reference values and sources for the rest wavelength ($\lambda_0$) of each line are given in Table~\ref{ism1}. For the DIBs with Gaussian fits we first 
       attempted fixing the FWHM using the reference value from the source. If that first attempt was successful, it is listed in the table as a single value. If
       we thought that the FWHM could be improved, we did a second attempt leaving it as a free parameter for each fit. Those cases are listed with a measurement and 
       an uncertainty in Table~\ref{ism1} and those with the most significant changes also appear in Table~\ref{ism2}.
\end{itemize}

\begin{table}
\centerline{
\begin{tabular}{cccc}
\multicolumn{2}{c}{$\lambda_0$} & \multicolumn{2}{c}{FWHM}  \\
\multicolumn{2}{c}{(\AA)}       & \multicolumn{2}{c}{(\AA)} \\
old         & new               & old         & new         \\
\hline
4427.94     & 4428.42           & 24.15       & 17.30       \\
4761.12     & 4765.0            & 19.72       & 22.00       \\
\multicolumn{2}{c}{4779.69}     & 5.48        & 3.00        \\
\multicolumn{2}{c}{4963.85}     & 2.62        & 0.66        \\
\multicolumn{2}{c}{4984.59}     & 1.33        & 0.55        \\
\multicolumn{2}{c}{5236.29}     & 2.27        & 1.56        \\
5449.83     & 5451.1            & \multicolumn{2}{c}{14.06} \\
\multicolumn{2}{c}{5487.23}     & 6.63        & 4.78        \\
\multicolumn{2}{c}{5494.29}     & 1.90        & 0.71        \\
6010.75     & 6010.43           & \multicolumn{2}{c}{3.27}  \\
6283.84     & 6284.36           & \multicolumn{2}{c}{4.77}  \\
6590.42     & 6591.5            & 7.53        & 12.82       \\
8621.20     & 8620.65           & 1.86        & 4.69        \\
\hline
\end{tabular}}
\caption{DIBs for which the measured values of $\lambda_0$ and/or FWHM are significantly different for \LSXI\ and \LSXII\ when compared with the reference values.
A single value indicates no apparent change. The number of significant digits in $\lambda_0$ reflects the precision of the value, which is worse for weak/broad DIBs
than for strong/narrow ones.
The values listed here originate from both the high- and intermediate-resolution spectroscopy (see Table~\ref{ism1}) but the latter is not used for the narrow DIBs 
to avoid resolution issues.
}
\label{ism2}
\end{table}

The EWs obtained from the GTC spectra for the three targets are given in Table~\ref{ism3}. We followed the same procedure as for the WHT and high-resolution
spectroscopy with only a few differences:

\input{ism3}

\begin{itemize}
 \item We only measured those lines where additional information could be added to the one in Table~\ref{ism1}. 
 \item For \LSXI~B we extracted the absorption lines after subtracting a model TLUSTY SED (see below). This allowed us to measure \CaII{3968.468} for that sightline (but
       not for the other two).
 \item The three CN lines at 3873.999+74.607+75.760 could not be individually resolved so we give the total EW.
 \item For \CHp{3957.692} and \CaI{4226.7275} we would not obtain an EW for \LSXI~B due to the poor S/N. 
 \item A single measurement flag (Mf) is used to indicate either a Gaussian fit (F) or numerical integration (I) was used to obtain the EW.
 \item In most cases the uncertainties for \LSXI\ and \LSXII\ are larger than in Table~\ref{ism1} despite the larger aperture of GTC compared to the other telescopes. 
       This is caused by the lower spectral resolution, fewer epochs, and shorter exposure times.
\end{itemize}

\section{\LSXI~B}

\begin{figure*}
\centerline{\includegraphics[width=\linewidth]{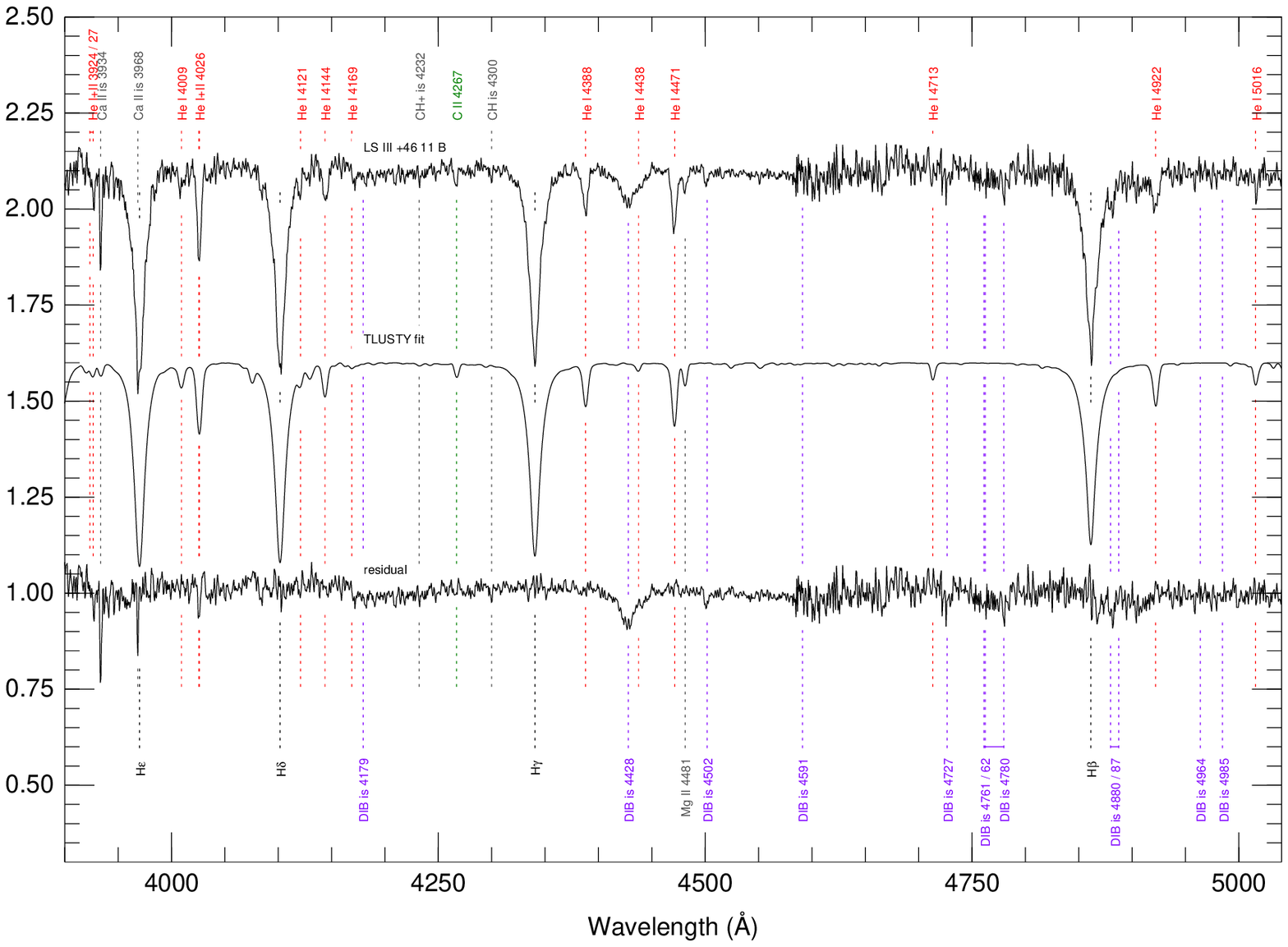}}
\caption{[top] GTC combined spectrum of \LSXI~B. [middle] Best TLUSTY fit for \LSXI~B. [bottom] Residual obtained by dividing the top spectrum by the middle fit.}
\label{LSXIB}
\end{figure*}

$\,\!$\indent The blue-violet section of the GTC combined spectrum of \LSXI~B is shown at the top of Fig.~\ref{LSXIB}. The jump in S/N at 4590~\AA\ is caused by the 
use of two gratings, R2500U and R2500V, to the left and right, respectively, with different exposure times. 

As we did in Paper I for \LSXI\ and \LSXII, we used MGB \citep{Maizetal12} to derive the stellar properties of \LSXI~B. 
We built a classification grid based not on observed spectra but on the TLUSTY grids of \citet{LanzHube03,LanzHube07} using
their SYNSPEC output. The grid uses \teff\ (15\,000-55\,000~K) as a horizontal coordinate and \logg\ (1.75-4.75 cgs) as vertical coordinate and, as before, MGB allows 
for the variation of the rotation speed \vsini.

MGB yields a good fit to the observed spectrum with $\teff = 16\,000\pm 1000$~K, $\logg = 4.00\pm 0.25$~cgs, and $\vsini = 175\pm 25$~km/s. \LSXI~B appears to be
a normal main-sequence mid-B star\footnote{
Comparison with our original GOSSS standard grid extended to B stars using MGB yields a B4~V spectral type \citep{Maizetal15b}.
} with a moderately high rotation speed. The value of \logg\ is slightly lower than the expected one for a $\sim$4.5~M$_\odot$ star of 
the young age of \BXC\ but the discrepancy is of only 1~sigma. Note that the estimated uncertainties are obtained by eye comparing different models in the
grid i.e. they are not formal uncertainties derived by e.g. $\chi^2$ fitting. The best TLUSTY fit is shown in the middle of Fig.~\ref{LSXIB} and the residual of the
fit at the bottom. The residual yields the ISM spectrum used to measure the EWs of its absorption lines. In particular, note how the residual has 
subtracted H$\varepsilon$ correctly, leaving the narrow ISM \CaII{3968.468} line clearly visible.

\begin{table}
\centerline{
\begin{tabular}{lr@{$\pm$}lr@{$\pm$}lr@{$\pm$}l}
Quantity           & \multicolumn{2}{c}{Baseline}   & \multicolumn{2}{c}{Alt. LC}    & \multicolumn{2}{c}{Alt. \teff} \\
\hline
\teff\ (K)         & \multicolumn{2}{c}{16\,000}    & \multicolumn{2}{c}{16\,000}    & \multicolumn{2}{c}{17\,000}    \\
luminosity class   & \multicolumn{2}{c}{5.5}        & \multicolumn{2}{c}{5.0}        & \multicolumn{2}{c}{5.5}        \\
\chir              & \multicolumn{2}{c}{0.50}       & \multicolumn{2}{c}{0.48}       & \multicolumn{2}{c}{0.51}       \\
\rv                & \multicolumn{2}{c}{3.34}       & \multicolumn{2}{c}{3.34}       & \multicolumn{2}{c}{3.34}       \\
\ebv\              &  1.590  & 0.019                &  1.590  & 0.019                &  1.606  & 0.019                \\
$\;\;$ (mag)       & \multicolumn{2}{c}{}           & \multicolumn{2}{c}{}           & \multicolumn{2}{c}{}           \\
\AV\  (mag)        &  5.324  & 0.064                &  5.323  & 0.064                &  5.376  & 0.063                \\
$V_{J,0}$ (mag)    & 10.668  & 0.040                & 10.669  & 0.040                & 10.620  & 0.040                \\
\logd\ (pc)        &  3.197  & 0.008                &  3.335  & 0.008                &  3.223  & 0.008                \\
\hline
\end{tabular}
}
\caption{Results of the CHORIZOS fits for \LSXI~B.}
\label{chorizos_output}
\end{table}

We have also done a CHORIZOS \citep{Maiz04c} analysis of \LSXI~B, similar to those in Paper I for the O stars but with some differences:

\begin{itemize}
 \item As input photometry we used 2MASS ($JHK_{\mathrm{S}}$, \citealt{Skruetal06}) and IPHAS ($ri$, \citealt{Bareetal14}), as there is no other high-quality 
       optical data available. As it happened for \LSXI\ with $K_{\mathrm{S}}$, there is no good detection of \LSXI~B in the 2MASS $J$ photometry. We downloaded the
       image from the archive and we performed a differential photometry analysis similar to the one we did for \LSXI\ in Paper I. That yielded a $J$ magnitude of
       $12.552\pm0.050$ for \LSXI~B.
 \item We used the same Milky Way grid \citep{Maiz13a} and extinction laws \citep{Maiz13b,Maizetal14a} as in Paper I. 
 \item For the baseline CHORIZOS run we fixed \teff\ to 16\,000~K (see above), the photometric luminosity class (LC) to 5.5 (ZAMS, since the star is expected to be 
       $\sim$2~Ma, young for the lifetime of a mid-B star in the main sequence), and the extinction type (\rv) to 3.34 (an average of the two very similar values 
       measured in Paper I for \LSXI\ and \LSXII). The amount of extinction [\ebv] and logarithmic distance (\logd) were left as free parameters. 
 \item We did two additional runs, one changing LC to 5.0 (average main sequence) and another changing \teff\ to 17\,000 K to quantify the effect of the parameter
       uncertainty in the output.
\end{itemize}

The CHORIZOS results are shown in Table~\ref{chorizos_output}:

\begin{itemize}
 \item The three runs give similar results for \chir\ and \ebv\ and only differ significantly
       in \logd. This happens because the optical+NIR colors to the right of the Balmer jump for early/mid B stars are nearly independent of luminosity
       and only weakly dependent on \teff. 
 \item The values of \chir\ indicate that the fit is very good in all cases (due to the reasons in the previous point they are all expected to be similar). This is in
       principle a sign of the validity of the extinction laws and the \rv\ value but it is only a weak sign, as the SED is not probed for $\lambda < 6000$~\AA.
 \item The \LSXI~B amount of extinction is intermediate between those of \LSXI\ and \LSXII, but significantly closer to that of \LSXI\ (which is a shorter distance away
       in the plane of sky). 
 \item If \LSXI~B is close to the ZAMS, \logd\ must be close to 3.20. Allowing for an older age places the system slightly farther away but not as far as the preferred
       value of 3.40-3.45 derived in Paper I, which required that \LSXII\ is a hidden binary. A \logd\ close to 3.20 poses no problem for a single \LSXII\ but it implies
       that, if all the stars are at the same distance, the two spectroscopic components of \LSXI\ would have similar visual luminosities as \LSXII, which contradicts their
       supergiant spectroscopic classification. An alternative, already hinted in Paper I, would imply that \HeII{4685.71} for the earliest O stars is a measurement of wind
       strength but not of luminosity. 
\end{itemize}

\section{ISM results}

\begin{figure}
\centerline{\includegraphics[width=\linewidth]{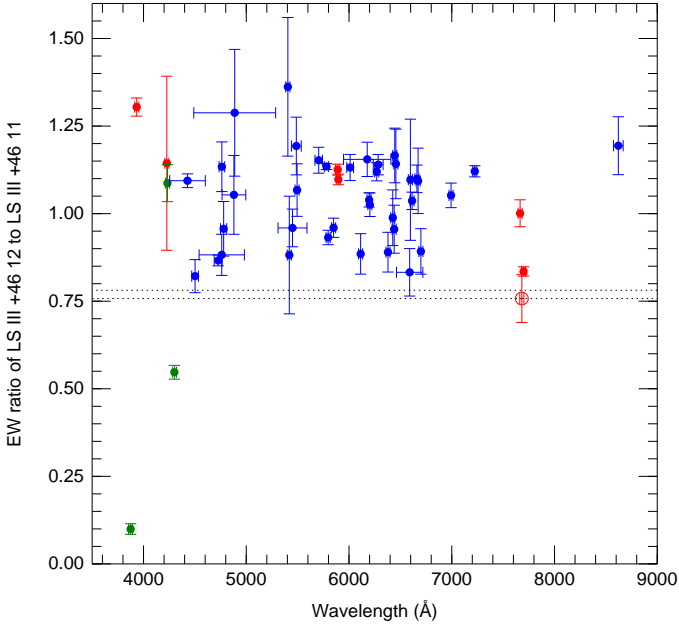}}
\caption{Equivalent-width ratios between \LSXII\ and \LSXI\ for the cases in Table~\ref{ism1} where measurements for both stars could be obtained. Red is used for atomic
lines, green for molecular lines, and blue for DIBs. All points show the EW uncorrected for saturation (which should be a good approximation in all cases except for the 
intense atomic lines) except for \KI{7664.911+7698.974}. In that case, besides the uncorrected values we show the corrected one (as an unfilled symbol) calculated using the 
kinematic decomposition described in the text. The $x$ value in each case is $\lambda_0$, the length of the horizontal error bars is proportional (with a 
threshold for the narrowest ones) to the FWHM of the line, and the vertical error bars show the uncertainty in the measurement. The two dotted lines are the ratios for
\ebv\ and \AV.}
\label{ewr}
\end{figure}

$\,\!$ \indent We analyze the ISM lines in three steps. We first concentrate on the velocities in Table~\ref{ism1} and the EWs in Tables~\ref{ism1}~and~\ref{ism3}:

\begin{itemize}
 \item All the EWs with values in both Tables~\ref{ism1}~and~\ref{ism3} are consistent within the errors, so from now on we will refer only to the most precise value of the
       two in each case.
 \item All the atomic and molecular lines for \LSXI\ have velocities between -20.6 and -13.8 km/s. \CHp{4232.548}, \CH{4300.313}, and \KI{7664.911+7698.974} for \LSXII\ 
       also have similar velocities. However, the \CaII{3933.663} and \NaI{5889.951+5895.924} velocities for that star are significantly larger, especially in the first case. 
       The cause is explained below.
 \item The two R(0) lines of CH+ are easily detected in the NoMaDS spectra of \LSXI\ but the R(1) and Q(1) lines are not, possibly due to an insufficient S/N \citep{Okaetal13}.
 \item The majority of the DIBs for both stars have velocities close to those of the atomic and molecular lines, indicating that the DIBs originate at the same clouds
       (from one point of view) or that we have a good knowledge of their central wavelengths (from another point of view). The most clear exceptions are listed in 
       Table~\ref{ism2}, where we computed the new values of $\lambda_0$ assuming that the DIBs originated at the same velocity as the atomic and molecular lines. The 
       majority of the changes listed there correspond to values measured by \citet{Maizetal14b}, who used only intermediate-resolution spectroscopy (which
       biases FWHM towards larger values) and combined profiles from different stars (which can also broaden the FWHM as well as introduce biases in $\lambda_0$).
       Among those cases we find the well known DIB with $\lambda\approx 4428$~\AA, which is very strong but broad and contaminated by weak stellar lines: the value found here 
       is $\approx$0.5~\AA\
       towards the blue compared with the one in \citet{Maizetal14b}. The 8621~\AA~DIB was not included on the \citet{Hobbetal08} or \citet{Maizetal14b} lists so the
       reference values listed in Tables~\ref{ism1}~and~\ref{ism2} are from \citet{JennDese94}. This DIB has received recent attention due to its inclusion in 
       large-scale surveys such as RAVE and Gaia, so it is important to know its intrinsic characteristics with accuracy. Our values for $\lambda_0$ and FWHM are 
       significantly different from the reference ones but we should note that more recent studies (e.g. \citealt{Munaetal08,Hobbetal09,Kosetal13}) have also pointed
       towards changes in the same direction (lower central wavelengths and broader profiles). 
 \item The EWs for the \LSXI~B lines have, in general, relatively large errors. However, they are all consistent with having the same values as for \LSXI, pointing towards 
       similar properties for both sightlines (as also indicated by the similar extinction values). 
 \item The EW ratios between \LSXII\ and \LSXI\ are plotted in Fig.~\ref{ewr}, along with the corresponding ratios for the dust measurements, 0.76 and 0.78 for \ebv\ and \AV,
       respectively (as previously noted, the two stars show similar values of \rv, hence the similar values for the two ratios).
 \item The behavior for the EW ratios of the molecular lines in Fig.~\ref{ewr} is very different. \CHp{4232.548} has a value of 1.09, clearly above the values for dust and 
       similar to the values for the DIBs (see below). On the other hand, the value for \CH{4300.313}, 0.55, is clearly below the dust values and the one for 
       \CN{3873.999+74.607+75.760}, 0.10, is much lower.
 \item Based on the previous points, if we model the ISM towards \BXC\ as a common cloud (cloud $\sigma$) that affects \LSXI\ and \LSXII\ similarly and a second cloud (cloud 
       $\zeta$) that affects \LSXI\ almost exclusively\footnote{
       The reason for the cloud nomenclature is explained later on.
       }, cloud $\zeta$ would have little or nothing CH+ but a significant amount of dust, an even larger (in relative terms) amount of CH, and more CN
       by an order of magnitude\footnote{We should think of these two clouds not as two exclusive physical entities. Indeed, as described below, cloud $\zeta$ appears to be 
       associated with the cluster while cloud $\sigma$ appears to have two components, one associated with the cluster and one located in the
       path between \BXC\ and the Sun.}. Given that CH traces gas denser than CH+ \citep{Smoketal14}, cloud $\zeta$ should be denser than cloud $\sigma$. 
 \item A similar comparison for atomic EW ratios is less straightforward because of saturation effects associated with the larger optical depths. The ratio for \KI{7698.974}, 
       which is optically thinner than the Na\,{\sc i} and Ca\,{\sc ii} lines, is 0.85, which is relatively close to the dust ratios. See below for a more detailed analysis taking 
       into account the kinematics and saturation.
 \item There is a significant scatter in the vertical axis of Fig.~\ref{ewr} for the DIBs, as expected, since the correlations between DIBs and extinction are good but
       not perfect. However, what is more surprising is that for all cases the EW ratio is higher than for \ebv\ or \AV. If we follow the two-cloud hypothesis previously 
       described, that would mean that cloud $\zeta$ is depleted in DIBs with respect to cloud $\sigma$.
 \item Within the scatter shown in Fig.~\ref{ewr}, $\zeta$-type DIBs (e.g. \DIB{5797.06} and \DIB{5849.81}, \citealt{Kreletal97}) have values of $\approx$1.0 or below while
       $\sigma$-type DIBs (e.g. \DIB{5780.48}) have values of $\approx$1.1 or above. This is another indication that cloud $\zeta$ should be interpreted as being denser than 
       cloud $\sigma$.
 \item \citet{McCaetal10} discovered that the 6195.98~\AA\ and 6613.62~\AA~DIBs are nearly perfectly correlated. \LSXI\ and \LSXII\ are more extinguished than the majority of 
       stars in their sample but the strong correlation still appears to hold. The EW ratios for those two lines in Fig.~\ref{ewr} are 1.037$\pm$0.020 and 1.012$\pm$0.014 i.e. 
       one sigma away from each other, and the ratios of one line to another are 4.1-4.2, very close to the 3.96 value determined by \citet{McCaetal10}. Note, however, that 
       a nearly perfect correlation does not necessarily imply a same carrier \citep{Okaetal13}.
\end{itemize}

Second, we qualitatively discuss the existence of different kinematic components seen in some of the ISM lines:

\begin{itemize}
 \item There are two clear kinematic components in \KI{7664.911+7698.974} for \LSXI\ and \LSXII\ in the CAF\'E-BEANS data. A stronger one at -19.34$\pm$0.01~km/s for \LSXI\
       (-20.96$\pm$0.02~km/s for \LSXII) and a weaker one at -8.00$\pm$0.02 for \LSXI\ (-6.43$\pm$0.04~km/s for \LSXII). The larger velocity difference between components for
       \LSXII\ manifests itself in the height of the central peak in the top two spectra of Fig.~\ref{kin}.
 \item The strong component has a velocity consistent with that of \LSXI\ measured in Paper I and is relatively close to that of \LSXII. The weak component, on the other hand, 
       has a velocity intermediate between those of the stars and that of the Sun. We adopt as our model that the strong component originates in a cloud associated with \BXC\ and 
       that the weak one originates in a cloud in the path between the cluster and the Sun. In terms of the $\sigma$ and $\zeta$ clouds defined before, the strong component 
       would contribute to both clouds $\sigma$ and $\zeta$ while the weak one would contribute only to $\sigma$. Or, putting it in another way, the ISM common to both sightlines 
       could be described by a cloud $\sigma_1$ far away from the cluster (the weak kinematic component) and a cloud $\sigma_2$ associated with the cluster (part of the strong 
       kinematic component). In addition, the \LSXI\ sightline would have another (denser) cloud $\zeta$ associated with the cluster (i.e. with the same velocity as $\sigma_2$). 
       In the most simplified version of the model, $\sigma_1$ and $\sigma_2$ would produce the same effect in the spectra of \LSXI\ and \LSXII\ but we know that this cannot be 
       completely true, as some absorption lines have values above 1.0 in Table~\ref{ewr}. 
       This can be explained by $\sigma_2$ being longer along the \LSXII\ sightline than along the \LSXI\ one. Figure~\ref{model} represents this description.

\begin{figure}
\centerline{\includegraphics[width=\linewidth]{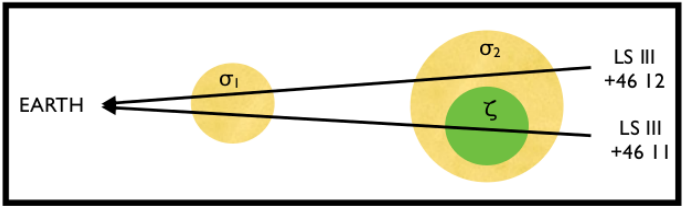}}
\caption{Model for the ISM clouds present in the \LSXI\ and \LSXII\ sightlines. Cloud $\sigma_1$ is of low density, not associated with the cluster, and affects both sightlines
similarly. Cloud $\sigma_2$ is also of low density, is the skin of the cloud associated with the cluster, and is longer along the \LSXII\ sightline. Cloud $\zeta$ is of high 
density, is the core of the cloud associated with the cluster, and affects \LSXI\ exclusively (or, at least, to a much larger degree than \LSXII).}
\label{model}
\end{figure}

 \item \NaI{5889.951+5895.924} is saturated for both stars but the profiles are consistent with the existence of the two same kinematic components as for K\,{\sc i}.
 \item \CaII{3933.663+3968.468} for \LSXI\ is also close to saturation and the only spectra with good S/N are from NoMaDS, so the resolution is lower than for K\,{\sc i} or
       Na\,{\sc i}. Also, for \LSXII\ the high-resolution spectra have low S/N in the 4000~\AA\ region so we only have GOSSS data there and only for \CaII{3933.663}, since it
       is not possible to separate \CaII{3968.468} from H$\varepsilon$ at the GOSSS lower resolution. In any case, the available data are consistent with the existence of the two 
       kinematic components detected in the K\,{\sc i} lines.
 \item The existence of two kinematic components can also be seen in the asymmetric profile of \CHp{4232.548} and \CH{4300.313} in \LSXI. 
 \item We do not detect two kinematic components in \CN{3873.999+3874.607+3875.760}, \CH{3886.410}, \CHp{3957.692}, or \CaI{4226.7275}, quite possibly due to an insufficient 
       S/N. However, their measured velocities in Table~\ref{ism1} indicate that they originate mostly in the strong K\,{\sc i} kinematic component. 
 \item In addition, there is a third high-velocity redshifted component detectable only for \LSXII\ in \NaI{5889.951+5895.924} and \CaII{3968.468}. The component is weak in
       Na\,{\sc i} but the high-resolution data allows us to measure a precise velocity of +196.04$\pm$0.22~km/s. The third component is much stronger in \CaII{3968.468}, making
       it easily detectable even at the GOSSS resolution (but note that we do not see it in either \LSXI\ or \LSXI~B). The strength of \CaII{3968.468} is a common feature in 
       high-velocity absorption components due to the lifting of the calcium depletion \citep{RoutSpit51,Walbetal02c} and is the reason why \CaII{3968.468} appears near the top
       of Fig.~\ref{ewr}. The high-velocity component is not seen in the K\,{\sc i} lines.
 \item We were unable to detect the two kinematic components in any of the DIBs. This is expected, given the spectral resolution of our data, the small velocity difference 
       between the two clouds, and the asymmetric and possibly variable profile of the narrow strong DIBs \citep{Galaetal08,Okaetal13}.
\end{itemize}

\begin{figure*}
\centerline{\includegraphics[width=0.14\linewidth]{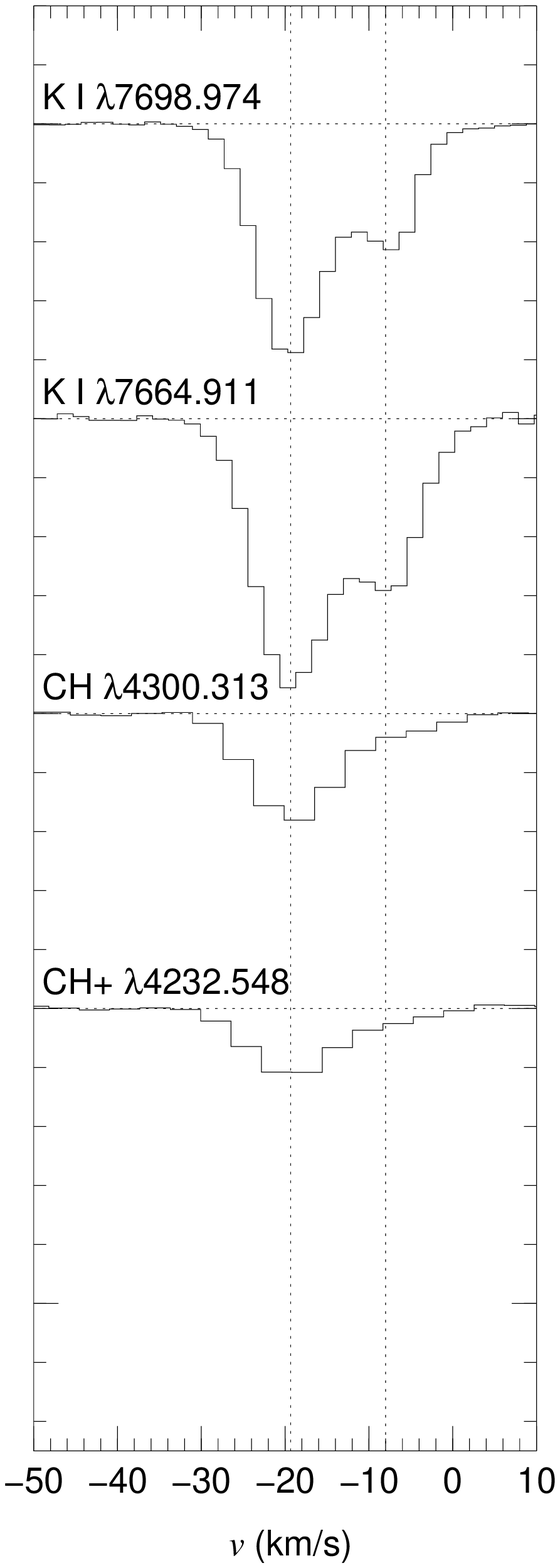} \
            \includegraphics[width=0.84\linewidth]{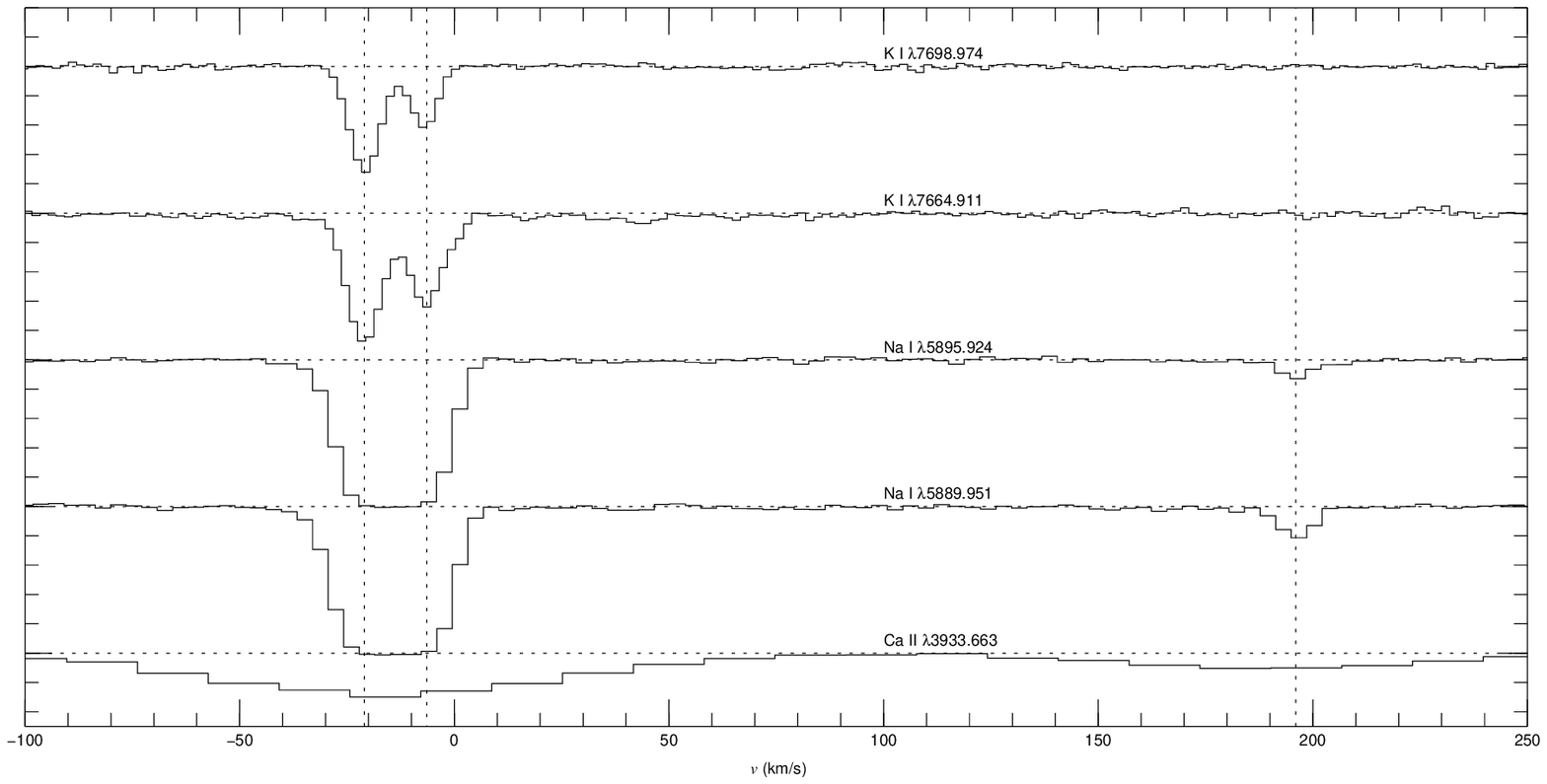}}
\caption{Line profiles with multiple components for \LSXI\ (left) and \LSXII\ (right). The K\,{\sc i} and Na\,{\sc i} spectra are from CAF\'E-BEANS, the CH and CH+ spectra from
NoMaDS, and the Ca\,{\sc ii} from GOSSS (hence, the significantly lower resolution, though the final spectrum has been drizzled from the original data to improve it.
The horizontal dotted lines indicate the normalized continuum level (the spectra are separated by one unit) and 
the vertical dotted lines indicate the velocities measured for K\,{\sc i} and for the third high-velocity component.}
\label{kin}
\end{figure*}

\begin{table*}
\centerline{
\begin{tabular}{lcccccc}
                            & \multicolumn{3}{c}{\LSXI}                        & \multicolumn{3}{c}{\LSXII}                       \\
Line(s)                     & St. comp.      & Wk. comp.      & Total          & St. comp.      & Wk. comp.      & Total          \\
\hline
\CN{3873.999+74.607+75.760} &                &                & 12.98$\pm$0.05 &                &                & 11.99$\pm$0.06 \\
\CHp{3957.692}              &                &                & 13.79$\pm$0.03 &                &                & 13.83$\pm$0.07 \\
\CaI{4226.7275}             &                &                & 10.70$\pm$0.04 &                &                & 10.76$\pm$0.08 \\
\CHp{4232.548}              & 13.73$\pm$0.02 & 13.25$\pm$0.20 & 13.85$\pm$0.02 &                &                & 13.89$\pm$0.05 \\
\CH{4300.313}               & 13.99$\pm$0.02 & 13.19$\pm$0.08 & 14.05$\pm$0.02 &                &                & 13.74$\pm$0.05 \\
\KI{7664.911+7698.974}      & 12.40$\pm$0.01 & 11.84$\pm$0.04 & 12.50$\pm$0.02 & 12.23$\pm$0.01 & 11.85$\pm$0.03 & 12.38$\pm$0.02 \\
\hline
\end{tabular}
}
\caption{Logarithm of the measured ISM column densities (in cm$^{-2}$). When high-resolution spectra are available, the column densities for the two components plus the total
value are given using the kinematic decomposition described in the text. In the other cases, we used the EWs from Tables~\ref{ism1}~and~\ref{ism3}. Since for \CHp{4232.548} and
\CH{4300.313} no high-resolution spectra were available for \LSXII\ and those lines have a slight saturation effect, we applied a correction to the EW based on the comparison 
between the intermediate-resolution and high-resolution results for those lines in \LSXI. The molecular oscillator strengths used 
are those of \citet{Smoketal14} while the ones for the atomic lines were obtained from the NIST web page {\tt http://physics.nist.gov/Pubs/AtSpec/table105.html}.}
\label{ism4}
\end{table*}

Third, we analyze the widths and column densities by kinematic component (Fig.~\ref{kin} and Table~\ref{ism4}) and calculate the column densities for the atomic and molecular 
species (whenever possible):

\begin{itemize}
 \item The values for the $b$ velocity parameter for the strong kinematic component are rather constant across different lines for the two stars, with values between 3.2 and
       3.6 km/s. The values for the weak component are similar but with a larger scatter, possibly due to the fitting noise induced by the presence of a stronger component.
       Given the large values of $b$, its most likely origin is ``turbulence'' (i.e. multiple components not detectable with the available resolution) rather than thermal motions.
 \item The weak component has very similar K\,{\sc i} column densities for both stars. That fact agrees with our model that it originates in a cloud ($\sigma_1$) not associated 
       with \BXC.
 \item When we compute the total column-density ratio between \LSXII\ and \LSXI\ for K\,{\sc i}, we find a value of 0.76$\pm$0.07, which is in agreement (within the errors) 
       with the ratios for \ebv\ and \AV\ (Fig.~\ref{ewr}). It appears that the lower value obtained directly from the EW ratios is indeed caused by a saturation effect. In other
       words, the comparison between the two sightlines provides a good correlation between the amounts of dust and K\,{\sc i} which is better than for any of the measured DIBs.
 \item Even though the uncertainty for the \CHp{4232.548} weak component is large, the ratio between the column densities of CH and CH+ is larger for the strong component than 
  for the weak one. pointing in favor of (some of) the strong component originating in a denser medium than the weak one. More specifically, if we include the values in the last 
       panel of Fig.~5 of \citet{Smoketal14}, [a] the weak component would have large values of $N$(CH) and $N$(CH+) but they would follow the average relation while [b] the
       strong component would be located at the top right corner (very large values of both column densities) with a larger than average $N$(CH)/$N$(CH+).
 \item We can only compare the total column densities of CN with those of CH and CH+. For \LSXII\, the three column densities fall reasonably well within the trends shown in 
       Fig.~5 of \citet{Smoketal14}. For \LSXI, CH and CN fall within their trend but both of them are above their expected values with respect to CH+. This indicates that
       all clouds (and especially cloud $\zeta$) are ``CN-like CH'' i.e. dense. If we use the \citet{Weseetal08} sample as a comparison, both stars would be among the ones with 
       the highest CH column densities.  \LSXI would also be in that category for CN but \LSXII\ would be among the targets with low column densities. 
\end{itemize}

\section{Discussion}

\input{ism5}

$\,\!$ \indent Our analysis of the ISM in front of \BXC\ points towards the existence of [a] one component ($\sigma_1$) located at a different velocity (and likely distance) than 
that of the cluster that affects both stars similarly and [b] a second component at a velocity similar to that of the cluster that affects \LSXI\ and \LSXI~B (as $\sigma_2+\zeta$)
more than \LSXII\ (as just $\sigma_2$). All clouds appear to be relatively dense but cloud $\zeta$ is denser than the average of clouds $\sigma_1$ and $\sigma_2$. This is shown by 
several indicators when comparing the equivalent widths different absorption lines for \LSXI\ (the star with the larger extinction) and \LSXII:

\begin{itemize}
 \item The EWs for \CH{4300.313} and \CN{3873.999+74.607+75.760} increase more than the extinction while \CHp{4232.548} increases less. 
 \item \KI{7664.911+7698.974} increases in a similar way as extinction (once saturation is taken into account). 
 \item All DIBs increase less than extinction (the skin effect, \citealt{Herb95}).
 \item $\zeta$-type DIBs (e.g. \DIB{5797.06}) increase more than $\sigma$-type DIBs (e.g. \DIB{5780.48}). 
       $\zeta$-type DIBs originate in different parts of the ISM, including UV-shielded, dense regions, while $\sigma$-type DIBs originate preferentially in UV-exposed, 
       thin regions of the ISM and are strongly depleted in dense regions \citep{Kreletal97,Camietal97}. This DIB property is the reason for the chosen cloud nomenclature.
 \item For the limited amount of lines with information for \LSXI~B, the ISM in front of that star appears to be more similar to \LSXI\ (to which is spatially closer)
       than to \LSXII.
\end{itemize}

Therefore, we predict that the remains of the molecular cloud from which \BXC\ formed are predominantly located towards the NW.
This description is also consistent with the WISE images, which show that the cluster has not evacuated a cavity yet and that there is an irregular dust distribution 
throughout the cluster. Hence, the surrounding ISM also points towards a young age for the cluster.

As previously mentioned, there are few sightlines in the literature 
with a large number of DIBs and even fewer for targets with extinction as high as these and with the ISM lines measured for the whole optical spectrum. 
The additional interest in this work is that we have a pair of sightlines with a common absorption component and another which affects only one star, with the second one being
[a] dense and [b] DIB-depleted. This provides the opportunity to rank the measured DIBs in a $\sigma-\zeta$ scale \citep{Kreletal97,Coxetal05} by sorting them by the
ratios of the equivalent widths plotted in Fig.~\ref{ewr}. We have done that in Table~\ref{ism5}:

\begin{itemize}
 \item The two DIBs that combine the properties of being apparently less affected by depletion in the dense ISM with being relatively narrow and strong are \DIB{4501.67} and 
       \DIB{4726.70}. Those would be good candidates to correlate better with extinction even when the ISM becomes very dense, a hypothesis that we plan to test in the future with
       other sightlines with high extinction (note that previous studies that included \DIB{4726.70} such as \citet{Puspetal13} tend to probe just the diffuse ISM). At longer 
       wavelengths the best candidate is still the $\zeta$-prototype, \DIB{5797.06}.
 \item At the other end of the scale, \DIB{8621.20}, present in the Gaia band, is a good example of a strong DIB that is expected to correlate poorly with extinction because
       it appears to be highly depleted in the dense ISM. A similar case is \DIB{5487.23}, which already had the worst correlation with extinction of the eight DIBs in 
       \citet{Frieetal11}.
 \item An interesting result is that for the pair \DIB{5404.56} + \DIB{5418.87}. They have relatively large errors in Table~\ref{ism5} but they are close in wavelength and in 
       opposite sides of the $\sigma-\zeta$ scale, making them good candidates to sample the diverse behavior of DIBs with density. Unfortunately, they are not very strong and they
       are placed at the wings of \HeII{5411.53}, which is strong for O stars.
 \item The order in Table~\ref{ism5} is in good agreement with the correlation with extinction for eight DIBs of \citet{Frieetal11}.
 \item There is no apparent correlation between the $\sigma-\zeta$ scale and the width of the DIBs: there are both narrow and broad DIBs at both ends of the scale.
\end{itemize}

Another interesting aspect of the intervening ISM is the location of \BXC\ well above the Galactic Plane. With a \logd\ of 3.40-3.45 and considering the Sun's own vertical distance
\citep{Maiz01a,Maizetal08a}, we find that \BXC\ is located almost 200~pc above the Galactic Plane, or about six times the scale height for OB stars. Even for a $\logd\sim 3.2$,
\BXC\ would be $\sim$150~pc above the Galactic Plane. 

We have convincingly shown that the amount of extinction is variable across the face of \BXC, with the value for \LSXI\ being
greater than that of \LSXII\ by 30\%, even though they are separated by a little over a pc in the plane of the sky. Also, the additional extinction
experienced by \LSXI\ appears to take place in a relatively dense cloud with properties different to the average of the ones that cause the rest of the
extinction along the line of sight (regarding e.g. DIBs). Yet, despite those differences, the measured \rv\ for both stars is essentially the same, 
indicating that there are no large differences in the average dust grain size for both sightlines. In other words, environments with different dust grain 
sizes are not required to produce differences in the observed DIBs.

\begin{acknowledgements}
We would like to thank Nolan R. Walborn and Ignacio Negueruela for useful comments on a previous version of this manuscript.
J.M.A. and A.S. acknowledge support from [a] the Spanish Government Ministerio de Econom{\'\i}a y Competitividad (MINECO) through grants 
AYA2010-15\,081, AYA2010-17\,631, and AYA2013-40\,611-P and [b] the Consejer{\'\i}a de Educaci{\'o}n of the Junta de Andaluc{\'\i}a through 
grant P08-TIC-4075. J.M.A. was also supported by the George P. and Cynthia Woods Mitchell Institute for Fundamental Physics and Astronomy.
He is grateful to the Department of Physics and Astronomy at Texas A\&M University for their hospitality during some of the time this work was 
carried out. 
R.H.B. acknowledges support from FONDECYT Project 1\,140\,076.
S.S.-D. acknowledges funding by [a] the Spanish Government Ministerio de Econom{\'\i}a y Competitividad (MINECO) through grants 
AYA2010-21\,697-C05-04, AYA2012-39\,364-C02-01, and Severo Ochoa SEV-2011-0187 and [b] the Canary Islands Government under grant PID2\,010\,119.
The data in this article were obtained with the 4.2~m William Hershel Telescope (WHT), the 10.4~m Gran Telescopio Canarias (GTC), and the 2.6~m Nordic 
Optical Telescope (NOT) at the Observatorio del Roque de los Muchachos (Spain); the 2.2~m and the 3.5~m telescopes at the Calar Alto Observatory (Spain); 
and the 9.2~m Hobby-Eberly Telescope (HET) at McDonald Observatory (U.S.A.).
\end{acknowledgements}

\bibliographystyle{aa}
\bibliography{general}

\end{document}

%% file: ism1.tex
\begin{table*}
{\scriptsize
\centerline{
\begin{tabular}{llr@{$\pm$}lr@{$\pm$}lr@{$\pm$}lr@{$\pm$}lr@{$\pm$}lr@{$\pm$}lccccc}
Type          & \multicolumn{1}{c}{$\lambda_0$} & \multicolumn{4}{c}{EW}                          & \multicolumn{4}{c}{FWHM}                        & \multicolumn{4}{c}{$v$}                         & Kf & \multicolumn{2}{c}{Mf} & Ref. & Notes \\
              & \multicolumn{1}{c}{(\AA)}       & \multicolumn{4}{c}{(m\AA)}                      & \multicolumn{4}{c}{(\AA)}                       & \multicolumn{4}{c}{km/s}                        &    &  &                     &      &       \\
              &                                 & \multicolumn{2}{c}{S1} & \multicolumn{2}{c}{S2} & \multicolumn{2}{c}{S1} & \multicolumn{2}{c}{S2} & \multicolumn{2}{c}{S1} & \multicolumn{2}{c}{S2} &    & S1 & S2                &      &       \\
\hline
CN            & 3873.999  &   25.2&  1.8 & \multicolumn{2}{c}{---} &  0.123&0.010              & \multicolumn{2}{c}{---}   &  -18.2& 0.4 & \multicolumn{2}{c}{---} & -  & HF & -  & 4 & -  \\
CN            & 3874.607  &   48.1&  1.9 & \multicolumn{2}{c}{---} &  0.144&0.007              & \multicolumn{2}{c}{---}   &  -17.8& 0.2 & \multicolumn{2}{c}{---} & -  & HF & -  & 4 & -  \\
CN            & 3875.760  &   14.4&  2.8 & \multicolumn{2}{c}{---} &  0.132&0.020              & \multicolumn{2}{c}{---}   &  -17.2& 0.7 & \multicolumn{2}{c}{---} & -  & HF & -  & 4 & -  \\
CH            & 3886.410  &   22.2&  2.8 & \multicolumn{2}{c}{---} &  0.135&0.019              & \multicolumn{2}{c}{---}   &  -20.6& 0.9 & \multicolumn{2}{c}{---} & -  & HF & -  & 4 & -  \\
Ca\,{\sc ii}  & 3933.663  &  390.8&  7.1 &  509.7&  4.0            & \multicolumn{2}{c}{---}   & \multicolumn{2}{c}{---}   &  -15.0& 0.5 &   35.0& 1.4             & M  & HI & GI & 5 & -  \\
CH+           & 3957.692  &   29.4&  1.8 & \multicolumn{2}{c}{---} &  0.188&0.013              & \multicolumn{2}{c}{---}   &  -18.8& 0.6 & \multicolumn{2}{c}{---} & -  & HF & -  & 4 & -  \\
Ca\,{\sc ii}  & 3968.468  &  290.4&  3.7 & \multicolumn{2}{c}{---} & \multicolumn{2}{c}{---}   & \multicolumn{2}{c}{---}   &  -15.0& 0.2 & \multicolumn{2}{c}{---} & M  & HI & -  & 5 & -  \\
Ca\,{\sc i}   & 4226.7275 &   13.9&  1.1 & \multicolumn{2}{c}{---} &  0.188&0.016              & \multicolumn{2}{c}{---}   &  -20.1& 0.8 & \multicolumn{2}{c}{---} & -  & HF & -  & 5 & -  \\
CH+           & 4232.548  &   46.0&  1.5 &   50.0&  1.8            & \multicolumn{2}{c}{---}   & \multicolumn{2}{c}{---}   &  -17.8& 0.2 &  -19.5& 1.7             & M  & HI & GI & 4 & -  \\
CH            & 4300.313  &   75.3&  1.3 &   41.2&  1.3            & \multicolumn{2}{c}{---}   & \multicolumn{2}{c}{---}   &  -17.3& 0.5 &  -23.1& 1.3             & M  & HI & GI & 4 & -  \\
DIB           & 4427.94   & 1641.4& 23.5 & 1795.6& 19.4            & 17.333&0.104              & 17.270&0.115              &   15.3& 3.1 &   14.9& 3.1             & -  & GF & GF & 3 & -  \\
DIB           & 4501.67   &  216.3&  6.9 &  177.7&  8.5            & \multicolumn{2}{c}{---}   & \multicolumn{2}{c}{---}   &   13.2& 1.9 &  -16.5& 1.7             & -  & GI & GI & 3 &  1 \\
DIB           & 4726.70   &  207.2&  2.3 &  179.5&  2.5            & \multicolumn{2}{c}{---}   & \multicolumn{2}{c}{---}   &    7.3& 2.1 &   -4.1& 1.6             & -  & GI & GI & 3 &  1 \\
DIB           & 4761.12   &  413.0& 13.1 &  364.4& 21.3            & 22.000&1.000              & 22.000&1.000              &  195.9&21.7 &  126.7&23.9             & -  & GF & GF & 3 &  6 \\
DIB           & 4762.36   &   87.4&  4.1 &   99.1&  4.1            & \multicolumn{2}{c}{ 2.69} & \multicolumn{2}{c}{ 2.69} &   -2.0& 4.2 &  -17.2& 3.7             & -  & GF & GF & 3 &  5 \\
DIB           & 4779.69   &   69.3&  3.9 &   66.3&  3.9            &  3.000&0.300              &  3.000&0.300              &   22.3& 6.0 &   -1.0& 6.3             & -  & GF & GF & 3 &  4 \\
DIB           & 4879.83   &  196.0& 12.5 &  206.5& 17.7            & \multicolumn{2}{c}{11.51} & \multicolumn{2}{c}{11.51} &  -13.6&13.6 & -116.7&24.9             & -  & GF & GF & 3 &  8 \\
DIB           & 4887.43   & 1128.5&136.4 & 1453.4&104.1            & \multicolumn{2}{c}{39.75} & \multicolumn{2}{c}{39.75} &  -13.6&13.6 & -116.7&24.9             & -  & GF & GF & 3 &  7 \\
DIB           & 4963.85   &   42.4&  1.5 & \multicolumn{2}{c}{---} &  0.659&0.027              & \multicolumn{2}{c}{---}   &  -14.0& 0.4 & \multicolumn{2}{c}{---} & -  & HF & -  & 3 & -  \\
DIB           & 4984.59   &   17.4&  1.2 & \multicolumn{2}{c}{---} &  0.549&0.032              & \multicolumn{2}{c}{---}   &   -6.7& 0.8 & \multicolumn{2}{c}{---} & -  & HF & -  & 3 & -  \\
DIB           & 5236.29   &   27.4&  2.6 & \multicolumn{2}{c}{---} &  1.561&0.099              & \multicolumn{2}{c}{---}   &  -15.5& 1.6 & \multicolumn{2}{c}{---} & -  & HF & -  & 3 & -  \\
DIB           & 5245.43   &  102.1&  5.2 & \multicolumn{2}{c}{---} & \multicolumn{2}{c}{ 7.26} & \multicolumn{2}{c}{---}   &   -2.6& 9.2 & \multicolumn{2}{c}{---} & -  & HF & -  & 3 & -  \\
DIB           & 5363.52   &   35.5&  3.9 & \multicolumn{2}{c}{---} &  2.228&0.224              & \multicolumn{2}{c}{---}   &   -7.4& 2.5 & \multicolumn{2}{c}{---} & -  & HF & -  & 3 & -  \\
DIB           & 5404.56   &   17.4&  1.1 &   23.7&  3.1            &  0.816&0.036              &  0.999&0.120              &  -20.5& 0.5 &  -15.8& 2.3             & -  & HF & HF & 1 & -  \\
DIB           & 5418.87   &   23.7&  0.9 &   20.9&  3.9            &  0.741&0.019              &  0.811&0.120              &  -17.3& 0.4 &  -14.5& 2.2             & -  & HF & HF & 1 & -  \\
DIB           & 5449.83   &  301.1& 11.7 &  288.8& 11.7            & \multicolumn{2}{c}{14.06} & \multicolumn{2}{c}{14.06} &  -46.5& 7.7 &  -54.9& 6.8             & -  & GF & GF & 3 & -  \\
DIB           & 5487.23   &  147.1&  6.9 &  175.5&  8.9            &  4.776&0.156              &  5.017&0.171              &   -1.7& 2.8 &  -17.2& 2.4             & -  & GF & GF & 3 & -  \\
DIB           & 5494.29   &   32.7&  0.8 &   34.9&  2.3            &  0.705&0.022              &  0.699&0.048              &  -28.7& 0.3 &  -25.7& 1.0             & -  & HF & HF & 3 & -  \\
DIB           & 5506.28   &   22.4&  0.9 & \multicolumn{2}{c}{---} & \multicolumn{2}{c}{ 1.30} & \multicolumn{2}{c}{---}   &  -20.9& 1.4 & \multicolumn{2}{c}{---} & -  & HF & -  & 1 & -  \\
DIB           & 5508.12   &   61.0&  3.1 & \multicolumn{2}{c}{---} &  1.899&0.042              & \multicolumn{2}{c}{---}   &  -21.3& 0.9 & \multicolumn{2}{c}{---} & -  & HF & -  & 1 & -  \\
DIB           & 5512.68   &   14.6&  0.5 & \multicolumn{2}{c}{---} & \multicolumn{2}{c}{ 0.71} & \multicolumn{2}{c}{---}   &  -15.9& 2.1 & \multicolumn{2}{c}{---} & -  & HF & -  & 1 & -  \\
DIB           & 5545.06   &   31.1&  2.6 & \multicolumn{2}{c}{---} &  0.921&0.051              & \multicolumn{2}{c}{---}   &  -16.6& 0.9 & \multicolumn{2}{c}{---} & -  & HF & -  & 1 & 10 \\
DIB           & 5705.08   &  146.3&  1.3 &  168.6&  5.2            & \multicolumn{2}{c}{ 3.70} & \multicolumn{2}{c}{ 3.70} &  -12.2& 0.7 &   -6.5& 2.8             & -  & HF & HF & 1 & -  \\
DIB           & 5780.48   &  496.8&  1.4 &  563.6&  3.9            & \multicolumn{2}{c}{---}   & \multicolumn{2}{c}{---}   &  -11.3& 0.2 &  -10.3& 0.5             & -  & HI & HI & 1 &  1 \\
DIB           & 5797.06   &  167.3&  1.4 &  155.9&  3.2            & \multicolumn{2}{c}{---}   & \multicolumn{2}{c}{---}   &  -12.7& 0.3 &  -12.7& 0.8             & -  & HI & HI & 1 &  1 \\
DIB           & 5849.81   &   61.9&  0.6 &   59.4&  1.6            & \multicolumn{2}{c}{ 0.82} & \multicolumn{2}{c}{ 0.82} &  -17.2& 0.2 &  -18.0& 0.6             & -  & HF & HF & 1 & -  \\
Na\,{\sc i}   & 5889.951  &  593.1&  3.3 &  667.6&  8.5            & \multicolumn{2}{c}{---}   & \multicolumn{2}{c}{---}   &  -14.2& 0.1 &    2.8& 1.2             & M  & HI & HI & 5 & -  \\
Na\,{\sc i}   & 5895.924  &  553.1&  3.4 &  607.1&  7.6            & \multicolumn{2}{c}{---}   & \multicolumn{2}{c}{---}   &  -14.3& 0.1 &   -2.1& 1.2             & M  & HI & HI & 5 & -  \\
DIB           & 6010.75   &  115.4&  1.4 &  130.6&  4.0            & \multicolumn{2}{c}{ 3.27} & \multicolumn{2}{c}{ 3.27} &  -31.9& 1.0 &  -34.6& 2.4             & -  & HF & HF & 1 & 11 \\
DIB           & 6113.18   &   20.0&  0.4 &   17.7&  1.1            & \multicolumn{2}{c}{ 0.68} & \multicolumn{2}{c}{ 0.68} &  -18.2& 0.4 &  -17.6& 1.1             & -  & HF & HF & 1 & -  \\
DIB           & 6139.98   &   10.5&  0.6 & \multicolumn{2}{c}{---} &  0.469&0.041              & \multicolumn{2}{c}{---}   &  -20.1& 0.9 & \multicolumn{2}{c}{---} & -  & HF & HF & 1 & -  \\
DIB           & 6177.30   & 1338.6& 54.8 & 1546.0& 17.0            & \multicolumn{2}{c}{23.06} & \multicolumn{2}{c}{23.06} & -105.9&12.3 & -103.6& 5.4             & -  & HF & HF & 2 &  2 \\
DIB           & 6195.98   &   56.3&  0.3 &   58.5&  1.1            & \multicolumn{2}{c}{ 0.42} & \multicolumn{2}{c}{ 0.42} &  -17.7& 0.1 &  -17.6& 0.4             & -  & HF & HF & 1 & 12 \\
DIB           & 6203.05   &   94.3&  1.0 &   96.7&  3.0            & \multicolumn{2}{c}{ 1.20} & \multicolumn{2}{c}{ 1.20} &  -18.5& 0.2 &  -17.7& 0.7             & -  & HF & HF & 1 &  3 \\
DIB           & 6269.85   &   86.1&  0.7 &   96.4&  2.1            & \multicolumn{2}{c}{ 1.18} & \multicolumn{2}{c}{ 1.18} &  -20.4& 0.2 &  -21.2& 0.6             & -  & HF & HF & 1 &  9 \\
DIB           & 6283.84   & 1037.0& 17.3 & 1182.0& 22.8            & \multicolumn{2}{c}{---}   & \multicolumn{2}{c}{---}   &    8.7& 1.8 &   12.8& 6.2             & -  & HI & HI & 1 &  1 \\
DIB           & 6379.32   &   70.0&  1.3 &   62.3&  3.8            & \multicolumn{2}{c}{ 0.59} & \multicolumn{2}{c}{ 0.59} &  -20.1& 0.1 &  -21.2& 0.7             & -  & HF & HF & 1 & -  \\
DIB           & 6425.66   &   17.1&  0.4 &   16.9&  1.3            & \multicolumn{2}{c}{ 0.75} & \multicolumn{2}{c}{ 0.75} &  -16.0& 0.4 &  -14.3& 1.5             & -  & HF & HF & 1 & -  \\
DIB           & 6439.48   &   18.0&  0.3 &   17.2&  1.2            & \multicolumn{2}{c}{ 0.75} & \multicolumn{2}{c}{ 0.75} &  -17.4& 0.6 &  -13.0& 1.4             & -  & HF & HF & 1 & -  \\
DIB           & 6445.28   &   26.5&  0.7 &   30.9&  1.9            & \multicolumn{2}{c}{---}   & \multicolumn{2}{c}{---}   &  -17.4& 0.3 &  -16.4& 1.3             & -  & HI & HI & 1 &  1 \\
DIB           & 6449.22   &   21.0&  0.8 & \multicolumn{2}{c}{---} &  0.823&0.045              & \multicolumn{2}{c}{---}   &  -18.5& 0.7 & \multicolumn{2}{c}{---} & -  & HF & HF & 1 & -  \\
DIB           & 6456.01   &   31.0&  0.6 &   35.4&  3.0            & \multicolumn{2}{c}{---}   & \multicolumn{2}{c}{---}   &  -18.2& 0.5 &  -20.6& 1.6             & -  & HI & HI & 1 &  1 \\
DIB           & 6520.62   &   19.0&  0.5 & \multicolumn{2}{c}{---} & \multicolumn{2}{c}{---}   & \multicolumn{2}{c}{---}   &  -15.4& 0.5 & \multicolumn{2}{c}{---} & -  & HI & -  & 1 &  1 \\
DIB           & 6590.42   &  412.5& 19.1 &  343.4& 23.0            & 12.823&0.248              & 10.402&0.714              &   14.8& 3.4 &   50.5& 7.8             & -  & HF & HF & 1 & -  \\
DIB           & 6597.31   &   12.4&  0.4 &   13.6&  2.1            & \multicolumn{2}{c}{ 0.54} & \multicolumn{2}{c}{ 0.54} &  -16.1& 0.5 &  -16.8& 2.1             & -  & HF & HF & 1 & -  \\
DIB           & 6613.62   &  237.3&  0.9 &  246.0&  5.8            & \multicolumn{2}{c}{---}   & \multicolumn{2}{c}{---}   &  -12.7& 0.1 &  -11.4& 0.5             & -  & HI & HI & 1 &  1 \\
DIB           & 6660.71   &   29.2&  0.3 &   32.1&  1.1            & \multicolumn{2}{c}{ 0.59} & \multicolumn{2}{c}{ 0.59} &  -19.9& 0.2 &  -18.3& 0.5             & -  & HF & HF & 1 & -  \\
DIB           & 6672.27   &   17.1&  0.5 &   18.7&  1.5            & \multicolumn{2}{c}{---}   & \multicolumn{2}{c}{---}   &  -17.9& 0.5 &  -22.0& 1.1             & -  & HI & HI & 1 &  1 \\
DIB           & 6699.32   &   24.1&  0.5 &   21.5&  1.5            & \multicolumn{2}{c}{ 0.64} & \multicolumn{2}{c}{ 0.64} &  -19.7& 0.3 &  -19.4& 1.1             & -  & HF & HF & 1 & 13 \\
DIB           & 6993.13   &   86.3&  0.4 &   90.8&  3.0            & \multicolumn{2}{c}{ 0.75} & \multicolumn{2}{c}{ 0.75} &  -18.1& 0.1 &  -18.5& 0.5             & -  & HF & HF & 1 & -  \\
DIB           & 7224.03   &  225.9&  1.3 &  253.2&  3.3            & \multicolumn{2}{c}{---}   & \multicolumn{2}{c}{---}   &  -14.0& 0.2 &  -13.3& 0.5             & -  & HI & HI & 1 &  1 \\
K\,{\sc i}    & 7664.911  &  390.1&  4.9 &  390.5& 14.2            & \multicolumn{2}{c}{---}   & \multicolumn{2}{c}{---}   &  -15.1& 0.1 &  -14.8& 0.4             & M  & HI & HI & 5 & -  \\
K\,{\sc i}    & 7698.974  &  299.8&  1.5 &  250.3&  3.8            & \multicolumn{2}{c}{---}   & \multicolumn{2}{c}{---}   &  -15.7& 0.1 &  -16.1& 0.2             & M  & HI & HI & 5 & -  \\
DIB           & 8621.20   &  305.3& 12.9 &  364.5& 20.0            &  4.688&0.127              &  3.684&0.236              &  -36.3& 2.0 &  -36.6& 2.4             & -  & HF & HF & 2 & -  \\
\hline
\end{tabular}}
\smallskip
References: 1: \citet{Hobbetal08}. 2: \citet{JennDese94}. 3: \citet{Maizetal14b}. 4: \citet{Smoketal14}. 5: \citet{vanH99}.\hfill$\;$\linebreak
Notes: 1: Asymmetric profile. 2: Broad and asymmetric but double fit not attempted. 3: Inside another broader unmeasured DIB, parabolic background used. 4: Multiple DIB wth 4761.12 \AA\ and 4762.36 \AA. 5: Multiple DIB wth 4761.12 \AA\ and 4779.69 \AA. 6: Multiple DIB wth 4762.36 \AA\ and 4779.69 \AA. 7: Multiple DIB wth 4879.83 \AA. 8: Multiple DIB wth 4887.43 \AA. 9: Possibly inside broader DIB. 10: Weaker DIB at 5546.08 \AA\ not fit. 11: Weaker DIB at 6004.89 \AA\ not fit. 12: Weaker DIB at 6194.74 \AA\ not fit. 13: Weaker DIB at 6702.02 \AA\ not fit.\hfill$\;$\linebreak
}
\caption{ISM lines measured for \LSXI\ (S1) and \LSXII\ (S2) in the WHT and high-resolution spectra. Empty values for the EW and $v$ indicate that no good-quality spectra were available for \LSXII. A single value 
         for the FWHM indicates that a reference value was used and left fixed, one with an uncertainty indicates that it was fitted in each spectrum, and an empty value that
         Gaussian fitting was not used or no good-quality spectra were available. Each line was fitted individually except for the two DIB groups at 4761.12+4762.36+4779.69 \AA\ 
         and 4879.83+4887.43 \AA\ \citep{Maizetal14b}, which were fitted simultaneously. Kf stands for kinematic multiplicity flag (one Gaussian component can be measured 
         at each of the two cloud velocities, see text). Mf is a measurement type flag with two components: [a] H indicates that the high-resolution data were used while 
         G indicates that the GOSSS WHT data were used and [b] F indicates that a Gaussian fit was used while I indicates that numerical integration was used instead.
         Ref. gives the reference used for the rest wavelength ($\lambda_0$) and for the FWHM (when fixed).}
\label{ism1}
\end{table*}

%% file: ism3.tex
\begin{table*}
\centerline{
\begin{tabular}{llr@{$\pm$}lr@{$\pm$}lr@{$\pm$}lc}
Type          & \multicolumn{1}{c}{$\lambda_0$} & \multicolumn{6}{c}{EW}                                                    & Mf \\
              & \multicolumn{1}{c}{(\AA)}       & \multicolumn{6}{c}{(m\AA)}                                                &    \\
              &                                 & \multicolumn{2}{c}{S1} & \multicolumn{2}{c}{S1B} & \multicolumn{2}{c}{S2} &    \\
\hline
CN            & 3873.999+74.607+75.760 &   87.4&  6.4 &  115.1& 49.0 &    8.7&  1.2 & F  \\
Ca\,{\sc ii}  & 3933.663               &  382.6&  6.5 &  396.1& 41.3 &  528.1&  8.4 & I  \\
CH+           & 3957.692               &   43.8&  9.4 & \multicolumn{2}{c}{---} &   31.7&  5.0 & F  \\
Ca\,{\sc ii}  & 3968.468               & \multicolumn{2}{c}{---} &  252.8& 31.8 & \multicolumn{2}{c}{---} & I  \\
Ca\,{\sc i}   & 4226.7275              &   15.2&  3.6 & \multicolumn{2}{c}{---} &   15.9&  3.1 & F  \\
CH+           & 4232.548               &   45.2&  2.5 &   43.0& 10.8 &   49.1&  3.6 & F  \\
CH            & 4300.313               &   74.9&  2.1 &   71.1&  4.4 &   43.8&  2.0 & F  \\
DIB           & 4427.94                & 1652.3& 13.6 & 1636.7& 75.7 & 1784.7& 12.2 & F  \\
DIB           & 4501.67                &  208.1&  4.9 &  186.3& 21.6 &  170.7& 11.5 & I  \\
DIB           & 5487.23                &  160.3& 23.9 &  151.4& 40.5 &  177.4& 16.2 & F  \\
DIB           & 5705.08                &  158.6& 13.2 &  166.3& 32.1 &  160.6& 10.0 & F  \\
DIB           & 5780.48                &  508.4&  8.8 &  526.4& 22.3 &  570.9&  7.4 & I  \\
DIB           & 5797.06                &  172.0&  8.8 &  201.6& 30.7 &  176.8& 31.6 & I  \\
Na\,{\sc i}   & 5889.951               &  593.8&  7.8 &  550.1& 47.7 &  655.9& 10.4 & I  \\
Na\,{\sc i}   & 5895.924               &  555.9& 20.2 &  559.4& 44.8 &  624.1& 23.1 & I  \\
\hline
\end{tabular}}
\smallskip
\caption{EWs measured for \LSXI\ (S1), \LSXI~B (S1B), and \LSXII\ (S2) in the GTC spectra. Empty values indicate that no measurement could be obtained.
         In the Mf column, F indicates that a Gaussian fit was used while I indicates that numerical integration was used instead.}
\label{ism3}
\end{table*}

%% file: ism5.tex
\begin{table}
\centerline{
\begin{tabular}{rcr@{$\pm$}lr}
\# & \multicolumn{1}{c}{$\lambda_0$} & \multicolumn{2}{c}{EW ratio} & \multicolumn{1}{c}{FWHM}  \\
   & \multicolumn{1}{c}{(\AA)}       & \multicolumn{2}{c}{}         & \multicolumn{1}{c}{(\AA)} \\
\hline
 1 & 4501.67 & 0.822&0.047 &  3.2 \\
 2 & 6590.42 & 0.832&0.068 & 12.8 \\
 3 & 4726.70 & 0.866&0.015 &  3.9 \\
 4 & 5418.87 & 0.882&0.168 &  0.7 \\
 5 & 4761.12 & 0.882&0.059 & 22.0 \\
 6 & 6113.18 & 0.885&0.058 &  0.7 \\
 7 & 6379.32 & 0.890&0.057 &  0.6 \\
 8 & 6699.32 & 0.892&0.065 &  0.6 \\
 9 & 5797.06 & 0.932&0.021 &  0.8 \\
10 & 6439.48 & 0.956&0.069 &  0.8 \\
11 & 4779.69 & 0.957&0.078 &  3.0 \\
12 & 5449.83 & 0.959&0.054 & 14.1 \\
13 & 5849.81 & 0.960&0.027 &  0.8 \\
14 & 6425.66 & 0.988&0.079 &  0.8 \\
15 & 6203.05 & 1.025&0.034 &  1.2 \\
16 & 6613.62 & 1.037&0.025 &  0.9 \\
17 & 6195.98 & 1.039&0.020 &  0.4 \\
18 & 6993.13 & 1.052&0.035 &  0.8 \\
19 & 4879.83 & 1.054&0.113 & 11.5 \\
20 & 5494.29 & 1.067&0.075 &  0.7 \\
21 & 6672.27 & 1.094&0.093 &  0.7 \\
22 & 4427.94 & 1.094&0.020 & 17.3 \\
23 & 6597.31 & 1.097&0.173 &  0.5 \\
24 & 6660.71 & 1.099&0.039 &  0.6 \\
25 & 6269.85 & 1.120&0.026 &  1.2 \\
26 & 7224.03 & 1.121&0.016 &  1.0 \\
27 & 6010.75 & 1.132&0.037 &  3.3 \\
28 & 4762.36 & 1.134&0.071 &  2.7 \\
29 & 5780.48 & 1.134&0.008 &  2.1 \\
30 & 6283.84 & 1.140&0.029 &  4.8 \\
31 & 6456.01 & 1.142&0.099 &  0.9 \\
32 & 5705.08 & 1.152&0.037 &  3.7 \\
33 & 6177.30 & 1.155&0.049 & 23.1 \\
34 & 6445.28 & 1.166&0.078 &  0.6 \\
35 & 5487.23 & 1.193&0.082 &  4.8 \\
36 & 8621.20 & 1.194&0.083 &  4.7 \\
37 & 4887.43 & 1.288&0.181 & 39.8 \\
38 & 5404.56 & 1.362&0.198 &  0.8 \\
\hline
\end{tabular}}
\smallskip
\caption{Sorted DIB equivalent-width ratios between \LSXII\ and \LSXI\ for the cases in Table~\ref{ism1} where measurements for both stars could be obtained.
         DIBs near the top should be less depleted in $\zeta$ clouds (dense), the prototype of such cases being \DIB{5797.06}. DIBs near the bottom should be more depleted
         in dense clouds and more prominent in $\sigma$ clouds (diffuse), the prototype of such cases being \DIB{5780.48}. Note that the uncertainties in
         some equivalent-width ratios are relatively high, so their placement in the sequence is uncertain. The FWHM is shown to differentiate narrow and broad DIBs.}
\label{ism5}
\end{table}

%% file: ms.bbl
\begin{thebibliography}{56}
\expandafter\ifx\csname natexlab\endcsname\relax\def\natexlab#1{#1}\fi

\bibitem[{Barentsen {et~al.}(2014)Barentsen, Farnhill, Drew,
  Gonz{\'a}lez-Solares, Greimel, Irwin, Miszalski, Ruhland, Groot, Mampaso,
  Sale, Henden, Aungwerojwit, Barlow, Carter, Corradi, Drake, Eisl{\"o}ffel,
  Fabregat, G{\"a}nsicke, Gentile~Fusillo, Greiss, Hales, Hodgkin, Huckvale,
  Irwin, King, Knigge, Kupfer, Lagadec, Lennon, Lewis, Mohr-Smith, Morris,
  Naylor, Parker, Phillipps, Pyrzas, Raddi, Roelofs, Rodr{\'{\i}}guez-Gil,
  Sabin, Scaringi, Steeghs, Suso, Tata, Unruh, van Roestel, Viironen, Vink,
  Walton, Wright, \& Zijlstra}]{Bareetal14}
Barentsen, G., Farnhill, H.~J., Drew, J.~E., {et~al.} 2014, MNRAS, 444, 3230

\bibitem[{Cami {et~al.}(1997)Cami, Sonnentrucker, Ehrenfreund, \&
  Foing}]{Camietal97}
Cami, J., Sonnentrucker, P., Ehrenfreund, P., \& Foing, B.~H. 1997, A\&A, 326,
  822

\bibitem[{Cox {et~al.}(2005)Cox, Kaper, Foing, \& Ehrenfreund}]{Coxetal05}
Cox, N.~L.~J., Kaper, L., Foing, B.~H., \& Ehrenfreund, P. 2005, A\&A, 438, 187

\bibitem[{Duke(1951)}]{Duke51}
Duke, D. 1951, ApJ, 113, 100

\bibitem[{Ehrenfreund {et~al.}(1995)Ehrenfreund, Foing, D'Hendecourt,
  Jenniskens, \& Desert}]{Ehreetal95}
Ehrenfreund, P., Foing, B.~H., D'Hendecourt, L., Jenniskens, P., \& Desert,
  F.~X. 1995, A\&A, 299, 213

\bibitem[{Friedman {et~al.}(2011)Friedman, York, McCall, Dahlstrom,
  Sonnentrucker, Welty, Drosback, Hobbs, Rachford, \& Snow}]{Frieetal11}
Friedman, S.~D., York, D.~G., McCall, B.~J., {et~al.} 2011, ApJ, 727, 33

\bibitem[{Galazutdinov {et~al.}(2008)Galazutdinov, LoCurto, \&
  Kre{\l}owski}]{Galaetal08}
Galazutdinov, G.~A., LoCurto, G., \& Kre{\l}owski, J. 2008, ApJ, 682, 1076

\bibitem[{Galazutdinov {et~al.}(2000)Galazutdinov, Musaev, Kre{\l}owski, \&
  Walker}]{Galaetal00}
Galazutdinov, G.~A., Musaev, F.~A., Kre{\l}owski, J., \& Walker, G.~A.~H. 2000,
  PASP, 112, 648

\bibitem[{Gardini {et~al.}(2013)Gardini, Ma{\'{\i}}z~Apell{\'a}niz, P{\'e}rez,
  Quesada, \& Funke}]{Gardetal13}
Gardini, A., Ma{\'{\i}}z~Apell{\'a}niz, J., P{\'e}rez, E., Quesada, J.~A., \&
  Funke, B. 2013, in Highlights of Spanish Astrophysics VII, 947--947

\bibitem[{Heger(1922)}]{Hege22}
Heger, M.~L. 1922, Lick Observatory Bulletin, 10, 146

\bibitem[{Herbig(1995)}]{Herb95}
Herbig, G.~H. 1995, ARA\&A, 33, 19

\bibitem[{Hobbs {et~al.}(2008)Hobbs, York, Snow, Oka, Thorburn, Bishof,
  Friedman, McCall, Rachford, Sonnentrucker, \& Welty}]{Hobbetal08}
Hobbs, L.~M., York, D.~G., Snow, T.~P., {et~al.} 2008, ApJ, 680, 1256

\bibitem[{Hobbs {et~al.}(2009)Hobbs, York, Thorburn, Snow, Bishof, Friedman,
  McCall, Oka, Rachford, Sonnentrucker, \& Welty}]{Hobbetal09}
Hobbs, L.~M., York, D.~G., Thorburn, J.~A., {et~al.} 2009, ApJ, 705, 32

\bibitem[{H{\o}g {et~al.}(2000)H{\o}g, Fabricius, Makarov, Urban, Corbin,
  Wycoff, Bastian, Schwekendiek, \& Wicenec}]{Hogetal00a}
H{\o}g, E., Fabricius, C., Makarov, V.~V., {et~al.} 2000, A\&A, 355, L27

\bibitem[{Jenniskens \& Desert(1994)}]{JennDese94}
Jenniskens, P. \& Desert, F.-X. 1994, A\&AS, 106, 39

\bibitem[{Ka{\'z}mierczak {et~al.}(2010)Ka{\'z}mierczak, Schmidt, Galazutdinov,
  Musaev, Betelesky, \& Kre{\l}owski}]{Kazmetal10}
Ka{\'z}mierczak, M., Schmidt, M.~R., Galazutdinov, G.~A., {et~al.} 2010, MNRAS,
  408, 1590

\bibitem[{Kos {et~al.}(2013)Kos, Zwitter, Grebel, Bienayme, Binney,
  Bland-Hawthorn, Freeman, Gibson, Gilmore, Kordopatis, Navarro, Parker, Reid,
  Seabroke, Siebert, Siviero, Steinmetz, Watson, \& Wyse}]{Kosetal13}
Kos, J., Zwitter, T., Grebel, E.~K., {et~al.} 2013, ApJ, 778, 86

\bibitem[{Krelowski {et~al.}(1997)Krelowski, Schmidt, \& Snow}]{Kreletal97}
Krelowski, J., Schmidt, M., \& Snow, T.~P. 1997, PASP, 109, 1135

\bibitem[{Lanz \& Hubeny(2003)}]{LanzHube03}
Lanz, T. \& Hubeny, I. 2003, ApJS, 146, 417

\bibitem[{Lanz \& Hubeny(2007)}]{LanzHube07}
Lanz, T. \& Hubeny, I. 2007, ApJS, 169, 83

\bibitem[{Maier {et~al.}(2011)Maier, Walker, Bohlender, Mazzotti, Raghunandan,
  Fulara, Garkusha, \& Nagy}]{Maieetal11}
Maier, J.~P., Walker, G.~A.~H., Bohlender, D.~A., {et~al.} 2011, ApJ, 726, 41

\bibitem[{Ma{\'{\i}}z~Apell{\'a}niz(2001)}]{Maiz01a}
Ma{\'{\i}}z~Apell{\'a}niz, J. 2001, AJ, 121, 2737

\bibitem[{Ma{\'{\i}}z~Apell{\'a}niz(2004)}]{Maiz04c}
Ma{\'{\i}}z~Apell{\'a}niz, J. 2004, PASP, 116, 859

\bibitem[{Ma{\'{\i}}z~Apell{\'a}niz(2013{\natexlab{a}})}]{Maiz13b}
Ma{\'{\i}}z~Apell{\'a}niz, J. 2013{\natexlab{a}}, in Highlights of Spanish
  Astrophysics VII, 583--589

\bibitem[{Ma{\'{\i}}z~Apell{\'a}niz(2013{\natexlab{b}})}]{Maiz13a}
Ma{\'{\i}}z~Apell{\'a}niz, J. 2013{\natexlab{b}}, in Highlights of Spanish
  Astrophysics VII, 657--657

\bibitem[{Ma{\'{\i}}z~Apell{\'a}niz
  {et~al.}(2015{\natexlab{a}})Ma{\'{\i}}z~Apell{\'a}niz, Alfaro, Arias,
  Barb{\'a}, Gamen, Herrero, Le{\~a}o, Marco, Negueruela,
  Sim{\'o}n-D{\'{\i}}az, Sota, \& Walborn}]{Maizetal15b}
Ma{\'{\i}}z~Apell{\'a}niz, J., Alfaro, E.~J., Arias, J.~I., {et~al.}
  2015{\natexlab{a}}, in Highlights of Spanish Astrophysics VIII, ed. A.~J.
  Cenarro, F.~Figueras, C.~Hern{\'a}ndez-Monteagudo, J.~Trujillo~Bueno, \&
  L.~Valdivielso, 603--603

\bibitem[{Ma{\'{\i}}z~Apell{\'a}niz {et~al.}(2008)Ma{\'{\i}}z~Apell{\'a}niz,
  Alfaro, \& Sota}]{Maizetal08a}
Ma{\'{\i}}z~Apell{\'a}niz, J., Alfaro, E.~J., \& Sota, A. 2008, arXiv:0804.2553

\bibitem[{Ma{\'{\i}}z~Apell{\'a}niz
  {et~al.}(2014{\natexlab{a}})Ma{\'{\i}}z~Apell{\'a}niz, Evans, Barb{\'a},
  Gr{\"a}fener, Bestenlehner, Crowther, Garc{\'{\i}}a, Herrero, Sana,
  Sim{\'o}n-D{\'{\i}}az, Taylor, van Loon, Vink, \& Walborn}]{Maizetal14a}
Ma{\'{\i}}z~Apell{\'a}niz, J., Evans, C.~J., Barb{\'a}, R.~H., {et~al.}
  2014{\natexlab{a}}, A\&A, 564, A63

\bibitem[{Ma{\'{\i}}z~Apell{\'a}niz
  {et~al.}(2015{\natexlab{b}})Ma{\'{\i}}z~Apell{\'a}niz, Negueruela, Barb{\'a},
  Walborn, Pellerin, Sim{\'o}n-D{\'{\i}}az, Sota, Marco, Alonso-Santiago,
  Sanchez~Bermudez, Gamen, \& Lorenzo}]{Maizetal15a}
Ma{\'{\i}}z~Apell{\'a}niz, J., Negueruela, I., Barb{\'a}, R.~H., {et~al.}
  2015{\natexlab{b}}, A\&A, 579, A108 (Paper I)

\bibitem[{Ma{\'{\i}}z~Apell{\'a}niz {et~al.}(2012)Ma{\'{\i}}z~Apell{\'a}niz,
  Pellerin, Barb{\'a}, Sim{\'o}n-D{\'{\i}}az, Alfaro, Morrell, Sota,
  Penad{\'e}s~Ordaz, \& Gallego~Calvente}]{Maizetal12}
Ma{\'{\i}}z~Apell{\'a}niz, J., Pellerin, A., Barb{\'a}, R.~H., {et~al.} 2012,
  in Astronomical Society of the Pacific Conference Series, Vol. 465,
  Astronomical Society of the Pacific Conference Series, ed. L.~Drissen,
  C.~Robert, N.~St-Louis, \& A.~F.~J. Moffat, 484

\bibitem[{Ma{\'{\i}}z~Apell{\'a}niz
  {et~al.}(2014{\natexlab{b}})Ma{\'{\i}}z~Apell{\'a}niz, Sota, Barb{\'a},
  Morrell, Pellerin, Alfaro, \& Sim{\'o}n-D{\'{\i}}az}]{Maizetal14b}
Ma{\'{\i}}z~Apell{\'a}niz, J., Sota, A., Barb{\'a}, R.~H., {et~al.}
  2014{\natexlab{b}}, in IAU Symposium, Vol. 297, IAU Symposium, ed. J.~Cami \&
  N.~L.~J. Cox, 117--120

\bibitem[{Ma{\'{\i}}z~Apell{\'a}niz {et~al.}(2011)Ma{\'{\i}}z~Apell{\'a}niz,
  Sota, Walborn, Alfaro, Barb{\'a}, Morrell, Gamen, \& Arias}]{Maizetal11}
Ma{\'{\i}}z~Apell{\'a}niz, J., Sota, A., Walborn, N.~R., {et~al.} 2011, in
  Highlights of Spanish Astrophysics VI, ed. {M.~R.~Zapatero Osorio, J.~Gorgas,
  J.~Ma{\'{\i}}z Apell{\'a}niz, J.~R.~Pardo, \& A.~Gil de Paz}, 467--472

\bibitem[{McCall {et~al.}(2010)McCall, Drosback, Thorburn, York, Friedman,
  Hobbs, Rachford, Snow, Sonnentrucker, \& Welty}]{McCaetal10}
McCall, B.~J., Drosback, M.~M., Thorburn, J.~A., {et~al.} 2010, ApJ, 708, 1628

\bibitem[{Merrill(1934)}]{Merr34}
Merrill, P.~W. 1934, PASP, 46, 206

\bibitem[{Munari {et~al.}(2008)Munari, Tomasella, Fiorucci, Bienaym{\'e},
  Binney, Bland-Hawthorn, Boeche, Campbell, Freeman, Gibson, Gilmore, Grebel,
  Helmi, Navarro, Parker, Seabroke, Siebert, Siviero, Steinmetz, Watson,
  Williams, Wyse, \& Zwitter}]{Munaetal08}
Munari, U., Tomasella, L., Fiorucci, M., {et~al.} 2008, A\&A, 488, 969

\bibitem[{Negueruela {et~al.}(2015)Negueruela, Ma{\'{\i}}z-Apell{\'a}niz,
  Sim{\'o}n-D{\'{\i}}az, Alfaro, Herrero, Alonso, Barb{\'a}, Lorenzo, Marco,
  Mongui{\'o}, Morrell, Pellerin, Sota, \& Walborn}]{Neguetal15}
Negueruela, I., Ma{\'{\i}}z-Apell{\'a}niz, J., Sim{\'o}n-D{\'{\i}}az, S.,
  {et~al.} 2015, in Highlights of Spanish Astrophysics VIII, ed. A.~J. Cenarro,
  F.~Figueras, C.~Hern{\'a}ndez-Monteagudo, J.~Trujillo~Bueno, \&
  L.~Valdivielso, 524--529

\bibitem[{Oka {et~al.}(2013)Oka, Welty, Johnson, York, Dahlstrom, \&
  Hobbs}]{Okaetal13}
Oka, T., Welty, D.~E., Johnson, S., {et~al.} 2013, ApJ, 773, 42

\bibitem[{Pellerin {et~al.}(2012)Pellerin, Ma{\'{\i}}z~Apell{\'a}niz,
  Sim{\'o}n-D{\'{\i}}az, \& Barb{\'a}}]{Pelletal12}
Pellerin, A., Ma{\'{\i}}z~Apell{\'a}niz, J., Sim{\'o}n-D{\'{\i}}az, S., \&
  Barb{\'a}, R.~H. 2012, in American Astronomical Society Meeting Abstracts,
  Vol. 219, American Astronomical Society Meeting Abstracts \#219, 224.03

\bibitem[{Penad{\'e}s~Ordaz {et~al.}(2013)Penad{\'e}s~Ordaz,
  Ma{\'{\i}}z~Apell{\'a}niz, \& Sota}]{Penaetal13}
Penad{\'e}s~Ordaz, M., Ma{\'{\i}}z~Apell{\'a}niz, J., \& Sota, A. 2013, in
  Highlights of Spanish Astrophysics VII, ed. J.~C. Guirado, L.~M. Lara,
  V.~Quilis, \& J.~Gorgas, 600--605

\bibitem[{Puspitarini {et~al.}(2013)Puspitarini, Lallement, \&
  Chen}]{Puspetal13}
Puspitarini, L., Lallement, R., \& Chen, H.-C. 2013, A\&A, 555, A25

\bibitem[{Raimond {et~al.}(2012)Raimond, Lallement, Vergely, Babusiaux, \&
  Eyer}]{Raimetal12}
Raimond, S., Lallement, R., Vergely, J.~L., Babusiaux, C., \& Eyer, L. 2012,
  A\&A, 544, A136

\bibitem[{Routly \& Spitzer(1951)}]{RoutSpit51}
Routly, P.~M. \& Spitzer, Jr., L. 1951, AJ, 56, 138

\bibitem[{Salama {et~al.}(2011)Salama, Galazutdinov, Kre{\l}owski, Biennier,
  Beletsky, \& Song}]{Salaetal11}
Salama, F., Galazutdinov, G.~A., Kre{\l}owski, J., {et~al.} 2011, ApJ, 728, 154

\bibitem[{Sim{\'o}n-D{\'{\i}}az {et~al.}(2011)Sim{\'o}n-D{\'{\i}}az, Castro,
  Garc{\'\i}a, \& Herrero}]{SimDetal11c}
Sim{\'o}n-D{\'{\i}}az, S., Castro, N., Garc{\'\i}a, M., \& Herrero, A. 2011, in
  IAU Symposium, Vol. 272, IAU Symposium, ed. C.~Neiner, G.~Wade, G.~Meynet, \&
  G.~Peters, 310--312

\bibitem[{Sim{\'o}n-D{\'{\i}}az {et~al.}(2015)Sim{\'o}n-D{\'{\i}}az,
  Negueruela, Ma{\'{\i}}z~Apell{\'a}niz, Castro, Herrero, Garcia,
  P{\'e}rez-Prieto, Caon, Alacid, Camacho, Dorda, Godart,
  Gonz{\'a}lez-Fern{\'a}ndez, Holgado, \& R{\"u}bke}]{SimDetal15b}
Sim{\'o}n-D{\'{\i}}az, S., Negueruela, I., Ma{\'{\i}}z~Apell{\'a}niz, J.,
  {et~al.} 2015, in Highlights of Spanish Astrophysics VIII, ed. A.~J. Cenarro,
  F.~Figueras, C.~Hern{\'a}ndez-Monteagudo, J.~Trujillo~Bueno, \&
  L.~Valdivielso, 576--581

\bibitem[{Skrutskie {et~al.}(2006)Skrutskie, Cutri, Stiening, Weinberg,
  Schneider, Carpenter, Beichman, Capps, Chester, Elias, Huchra, Liebert,
  Lonsdale, Monet, Price, Seitzer, Jarrett, Kirkpatrick, Gizis, Howard, Evans,
  Fowler, Fullmer, Hurt, Light, Kopan, Marsh, McCallon, Tam, Van~Dyk, \&
  Wheelock}]{Skruetal06}
Skrutskie, M.~F., Cutri, R.~M., Stiening, R., {et~al.} 2006, AJ, 131, 1163

\bibitem[{Smoker {et~al.}(2014)Smoker, Ledoux, Jehin, Keenan, Kennedy, Cabanac,
  \& Melo}]{Smoketal14}
Smoker, J., Ledoux, C., Jehin, E., {et~al.} 2014, MNRAS, 438, 1127

\bibitem[{Snow {et~al.}(2002)Snow, Zukowski, \& Massey}]{Snowetal02}
Snow, T.~P., Zukowski, D., \& Massey, P. 2002, ApJ, 578, 877

\bibitem[{Steglich {et~al.}(2011)Steglich, Bouwman, Huisken, \&
  Henning}]{Stegetal11}
Steglich, M., Bouwman, J., Huisken, F., \& Henning, T. 2011, ApJ, 742, 2

\bibitem[{Thorburn {et~al.}(2003)Thorburn, Hobbs, McCall, Oka, Welty, Friedman,
  Snow, Sonnentrucker, \& York}]{Thoretal03}
Thorburn, J.~A., Hobbs, L.~M., McCall, B.~J., {et~al.} 2003, ApJ, 584, 339

\bibitem[{Tuairisg {et~al.}(2000)Tuairisg, Cami, Foing, Sonnentrucker, \&
  Ehrenfreund}]{Tuaietal00}
Tuairisg, S.~{\'O}., Cami, J., Foing, B.~H., Sonnentrucker, P., \& Ehrenfreund,
  P. 2000, A\&AS, 142, 225

\bibitem[{van Hoof(1999)}]{vanH99}
van Hoof, P. 1999, UV, Optical \& Infrared Line List v2.04 (on line)
  (http://www.pa.uky.edu/$\sim$peter/atomic/index.html)

\bibitem[{van Loon {et~al.}(2013)van Loon, Bailey, Tatton,
  Ma{\'{\i}}z~Apell{\'a}niz, Crowther, de~Koter, Evans, H{\'e}nault-Brunet,
  Howarth, Richter, Sana, Sim{\'o}n-D{\'{\i}}az, Taylor, \&
  Walborn}]{vanLetal13}
van Loon, J.~T., Bailey, M., Tatton, B.~L., {et~al.} 2013, A\&A, 550, A108

\bibitem[{Vos {et~al.}(2011)Vos, Cox, Kaper, Spaans, \&
  Ehrenfreund}]{Vosetal11}
Vos, D.~A.~I., Cox, N.~L.~J., Kaper, L., Spaans, M., \& Ehrenfreund, P. 2011,
  A\&A, 533, A129

\bibitem[{Walborn {et~al.}(2002)Walborn, Danks, Vieira, \&
  Landsman}]{Walbetal02c}
Walborn, N.~R., Danks, A.~C., Vieira, G., \& Landsman, W.~B. 2002, ApJS, 140,
  407

\bibitem[{Weselak {et~al.}(2008)Weselak, Galazutdinov, Musaev, \&
  Kre{\l}owski}]{Weseetal08}
Weselak, T., Galazutdinov, G.~A., Musaev, F.~A., \& Kre{\l}owski, J. 2008,
  A\&A, 484, 381

\end{thebibliography}
